\documentclass[a4paper,12pt]{article}
\usepackage{amssymb,amsmath,mathrsfs, mathptmx}
\usepackage[usenames,dvipsnames]{xcolor}
\usepackage{hyperref, comment}
\usepackage{tensor}
\usepackage{graphicx}
\usepackage[left=1.2in,top=1in,right=1.2in,bottom=1in,headheight=0.8in,foot=0.5in]{geometry}
\usepackage[nodisplayskipstretch]{setspace} 
\usepackage[width=\textwidth]{caption}
\usepackage{pifont}% http://ctan.org/pkg/pifont
\usepackage{xcolor}
\usepackage{amsmath,latexsym}
\usepackage{outlines}
\usepackage[textsize=tiny]{todonotes}

\usepackage[style=ext-authoryear-comp,maxcitenames=2,uniquename=false,backend=biber,uniquelist=true,articlein=false]{biblatex}
\DeclareNameAlias{sortname}{last-first}

\usepackage[indentafter]{titlesec}
\titleformat{name=\section}{}{\thetitle.}{0.8em}{\centering\scshape}
\titleformat{name=\subsection}[runin]{}{\thetitle.}{0.5em}{\bfseries}[.]
\titleformat{name=\subsubsection}[runin]{}{\thetitle.}{0.5em}{\itshape}[.]
\titleformat{name=\paragraph,numberless}[runin]{}{}{0em}{}[.]
\titlespacing{\paragraph}{0em}{0em}{0.5em}
\titleformat{name=\subparagraph,numberless}[runin]{}{}{0em}{}[.]
\titlespacing{\subparagraph}{0em}{0em}{0.5em}

\newcommand{\nocontentsline}[3]{}
\newcommand{\tocless}[2]{\bgroup\let\addcontentsline=\nocontentsline#1{#2}\egroup}
\hypersetup{
	colorlinks=true,         
	linkcolor=Black,          
	citecolor=MidnightBlue,
	urlcolor=MidnightBlue            
 }

\title{Respecting Boundaries: Theoretical Equivalence and Structure Beyond Dynamics\vspace{-2em}\footnote{Published in \textit{Euro Jnl Phil Sci} 13, 47 (2023). https://doi.org/10.1007/s13194-023-00545-6.}}
\date{}

\makeatletter
\let\uppercasenonmath\@gobble

\begin{document}
%\setstretch{1.0}
\maketitle

\begin{center}
\author{William J.~Wolf\footnote{Faculty of Philosophy, University of Oxford, UK. Email: william.wolf@philosophy.ox.ac.uk} and James Read\footnote{Faculty of Philosophy, University of Oxford, UK. Email: james.read@philosophy.ox.ac.uk}}
\end{center}

\begin{abstract}
\noindent A standard line in the contemporary philosophical literature has it that physical theories are equivalent only when they agree on their empirical content, where this empirical content is often understood as being encoded in the equations of motion of those theories. In this article, we question whether it is indeed the case that the empirical content of a theory is exhausted by its equations of motion, showing that (for example) considerations of boundary conditions play a key role in the empirical equivalence (or otherwise) of theories. Having argued for this, we show that philosophical claims made by \textcite{Weatherall2016-WEAANG} that electromagnetism in the Faraday tensor formalism is equivalent to electromagnetism in the vector potential formalism, and by \textcite{Knox2011-KNONTA} that general relativity is equivalent to teleparallel gravity, can both be called into question. We then show that properly considering the role of boundary conditions in theory structure can potentially restore these claims of equivalence and close with some remarks on the pragmatics of adjudications on theory identity.
\end{abstract}
\tableofcontents
\setstretch{1.2}

% \begingroup
% \makeatletter
% \parskip\z@skip
% \tableofcontents
% \endgroup

\section{Introduction}

% Analyzes recent examples of supposed theoretical equivalence and shows that these adjudications are incorrect in terms of their statements on empirical equivalence. The source of these incorrect adjudications is traced to not accounting for boundary phenomena as part of the empirical content of the theories in question. Highlights the role and importance of boundary phenomena in discussions of theory structure and equivalence. Highlights how important view of theories is in these adjudications as well as how boundary phenomena has become more important in scientific practice.

\noindent Determining when two theories, models, or formulations of a theory are equivalent to one another (and in what sense) is a significant topic within the philosophy of science \parencite{Glymour1970-GLYTRA, Quine1975-QUIOEE, Weatherall2018-WEATEI}.
%\todo{Can we add some citations? Glymour, Barrett, Weatherall, etc.}\todoWill{Done}
The rationale underlying the attention which has been afforded to this issue presumably has to do with the idea that it is only through understanding these issues of equivalence that one can come to understand how a theory, model, or formulation comes to limn reality.
%because it adds conceptual clarity to the underlying mathematical structures used in the theories and to the claims that such descriptions make about the way the world is. Such further understanding has
Arguably, the quest for such understanding has also aided scientific progress in the past---examples include the equivalence of wave mechanics and matrix mechanics \parencite{vonNeumann+2018, Muller1997-MULTEM, Muller1997-MULTEM-2}, Feynmann and Swinger's approaches to quantum field theory \parencite{Dyson}, Lagrangian and Hamiltonian mechanics \parencite{Barrett2019-BAREAI-10, Curiel, North2009-NORTSO-14}, the AdS/CFT correspondence \parencite{Maldacena:1997re, deHaro2016-DEHCAO-2}, and many others.
%\todo{For each of these, can we add a citation? Muller 1997 SHPMP is good on the wave mechanics stuff.} \todoWill{Done}

%The most celebrated example of this is Von Neumann's demonstration of the equivalence between wave mechanics and matrix mechanics. This was particularly intriguing because it showed an intimate relationship between mathematical structures that, on the surface, seem completely disparate from each other, yet could still render the same empirical verdicts regarding quantum mechanical phenomena. It is not an exaggeration to say that this demonstration of equivalence has shaped our understanding of quantum mechanics and the way quantum mechanics is discussed and taught ever since. 

%Examples such as this have led the topic of theoretical equivalence to occupy notable corner within the philosophy of science landscape. 

The existing literature on theoretical equivalence is vast, and has focused on developing criteria for---and assessing the conditions under which---particular theories can be understood as being equivalent, as well as applying these criteria to specific examples in order to illuminate our understanding of particular theories and the interconnections that their structures may possess. In a recent discussion concerning the equivalence of Lagrangian and Hamiltonian mechanics, \textcite{Barrett2019-BAREAI-10} sketches an interesting connection between questions of theoretical equivalence and questions concerning the content or structure of a physical theory. While theoretical equivalence is certainly a significant topic within the philosophy of science, \textcite{VANFRAASSEN1986307} famously considers the question, `what is the content of a theory?', to be the central foundational question of philosophy of science. In identifying this relationship between questions of theoretical equivalence and the content of a theory, Barrett argues that whenever we commit to a method of identifying the content of a theory, we also necessarily commit to a standard of equivalence between theories. The converse also applies because
%Whenever we commit to a standard of equivalence between theories, we also necessarily commit to a method of identifying the content of a theory. 
when we commit to a particular standard of equivalence between theories, we are (for Barrett) also saying which features of our theories are significant or `contentful', as these are the very features that our assessment of equivalence will consider.

Within the philosophy of physics, philosophers typically (and justifiably) focus on the dynamical content encoded in the equations of motion as the relevant physical content of a theory. Often, this then (understandably) results in dynamical equivalence being taken to be a sufficient condition for empirical equivalence. Examples include \textcite{Weatherall2016-WEAANG} arguing for the theoretical equivalence of the electromagnetic field formulation and the gauge potential formulation of classical electromagnetism (EM) and \textcite{Knox2011-KNONTA} arguing for the theoretical equivalence of general relativity (GR) and the teleparallel equivalent of general relativity (TPG). In both cases, the authors maintain that the `contentful' features of the theories in question are fully captured in the theories' dynamics (or equivalently, their equations of motion), and in doing so adopt a standard of equivalence which holds that equivalent dynamics is sufficient for demonstrating empirical equivalence. For example, when discussing the empirical equivalence of the two formulations of EM, \textcite[p.~1078, our emphasis]{Weatherall2016-WEAANG} stipulates ``[G]iven a Faraday tensor $F_{ab}$ that satisfies Maxwell's equations [...] on both formulations, the empirical content of a model is \textit{exhausted} by its associated Faraday tensor. In this sense, the theories are empirically equivalent [...]''. Similarly, \textcite[p.~272]{Knox2011-KNONTA} indicates that ``the [TPG] Lagrangian above turns out to be identical, up to a divergence, to the Einstein-Hilbert Lagrangian in standard GR [...] the equivalence of the Lagrangians is enough to establish
empirical equivalence". As we shall see, while they do not \textit{explicitly} advocate a particular view of theory structure in their analyses, this standard of empirical equivalence (i.e., equivalent dynamics) nonetheless implies and is broadly consistent with a certain fairly typical version of the semantic view of scientific theories. This view, as usually articulated, holds that a theory's content is captured by models comprised of the right kinds of mathematical objects, where these objects obey some specified dynamics.

While this is certainly an understandable position given the prominence of dynamics in physical theories, recent decades have seen both philosophers and physicists investigating content that is not entirely determined by a theory's dynamics---in particular, the content inherent to describing isolated subsystems, their associated boundary conditions, and their relationship to their environments. Recently, philosophers have used isolated subsystems to investigate the empirical significance of gauge symmetries \parencite{Wallace2014-GREECO, Teh2016-TEHGGU, MurgueitioRamirezForthcoming-MURAGS-3, Gomes2021-GOMHAT, Wolf:2021ydy}, explored the important explanatory role that the boundary conditions associated with such isolated subsystems play in mathematical modeling \parencite{Bursten2021-BURTFO-13}, and considered the implications that boundary conditions might have for our conception of laws and the Humean mosaic \parencite{Travis}. Physicists have likewise focused on isolated subsystems and the boundary phenomena associated with them, as can be seen in prominent examples including edge modes in the quantum Hall effect \parencite{Wen:1995qn}, the study of black hole entropy \parencite{Gibbons:1976ue}, slip/no-slip boundary conditions in fluid flow \parencite{2005cond.mat..1557L}, and the AdS/CFT correspondence \parencite{Maldacena:1997re}.

Furthermore, when viewing the content of a physical theory as including the kinds of boundary content associated with isolated subsystems, it becomes clear that an analysis of empirical equivalence that relies upon only dynamics is deficient.\footnote{Throughout this article, we focus on the question of whether dynamical equivalence is sufficient for empirical equivalence by way of a study of boundary phenomena; we set aside other relevant factors in theory construction, e.g.~kinematical constraints (on which see \textcite{CurielKinematics, LinnemannReadKin}), the possibility of `schematising the observer' \textcite{curiel2020schematizing}, etc. We expect that taking into account these further considerations would only further bolster our point---\emph{viz}.,~that empirical content is not exhausted by dynamics.} In particular, in this paper we highlight how the analysis in the aforementioned examples from Knox and Weatherall does not account for such boundary phenomena;
%\todo{Wait, sorry, don't you mean `equivalent' here?} Failure to do so
%Incorporating consideration of boundaries 
doing so leads to a verdict that both pairs of theories, as presented by the authors, are in fact empirically (\emph{a fortiori} theoretically) \emph{inequivalent}. 
%This leads one to the conclusion either (i) that these theories truly are inequivalent, or (ii) that there is something unsatisfactory about the way in which the content of these theories was specified, thereby suggesting the following:
These results thereby invite the following conclusions:

%which leads to the conclusion that the theories considered are not equivalent 

%as articulated , are clearly empirically in-equivalent by exploring what these theories say about isolated subsystems and boundary phenomena. The impact is that it reveals that these adjudications of theoretical equivalence are, as argued, incorrect in light of further empirical content that has been identified as relevant to the respective theories and consequently motivates an examination of how the relevant content of a theory is identified. This leads to two payoffs. 

\begin{enumerate}
    \item  Adjudications of theoretical equivalence cannot be made independently of clearly committing oneself to particular judgments regarding a theory's relevant content. If one fails to account properly for the contentful features of a theory, one may be left with an adjudication of theoretical equivalence that is incorrect, a view of the theories' structure that is deficient, or both.
    \item The content of physical theories can extend beyond dynamics. Boundary phenomena, boundary conditions, and the modeling of subsystem-environment decompositions are relevant to questions concerning the content of physical theories, and likewise concerning theoretical equivalence, because these items are important for capturing the empirical content of physical theories. Indeed, some philosophers have begun to discuss boundary conditions alongside other elements that are typically invoked when specifying theoretical structure---see e.g.~\textcite{Wallace2014-GREECO, Teh2016-TEHGGU}. 
\end{enumerate}

\section{Views on Theoretical Equivalence}

\noindent Discussions of theoretical equivalence almost invariably begin with a notion of \textit{empirical equivalence}. If two theories disagree in terms of the empirical content associated with them, then no further analysis is necessary: they are inequivalent \emph{tout court}. The reason for this is that empirical goings-on are naturally regarded as supervening on physical goings-on. At a minimum, theories should necessarily have the same empirical content if they are to be considered equivalent. This means that two theories must have the same range of applicability regarding empirical scenarios they describe and provide indistinguishable predictions for the observational phenomena.
To be slightly more specific, we can understand that models $M$ of a theory $T$ will have empirical substructures, which can represent observable phenomena. Suppose, for every $M$ of $T$, there is an $M'$ of $T'$, where the empirical substructures of $M$ and $M'$ are isomorphic. Then, $T$ and $T'$ can be understood to be empirically equivalent \parencite{VanFraassenBas1980-VANTSI}. 

This is a fairly general way of stating what empirical equivalence amounts to. As we have seen above, showing that two theories possess equivalent dynamical content through their equations of motion is often taken to be sufficient to demonstrate empirical equivalence within the physical sciences. While we do not attempt to provide a fully exhaustive and all-encompassing definition of empirical equivalence (we can be pragmatic about this!---see \S\ref{Pragmatic}), one of the goals of this paper is to demonstrate clearly that within the physical sciences there is important content \textit{beyond} dynamics that should factor into our analyses of empirical equivalence. That is, dynamical equivalence alone is not sufficient to establish empirical equivalence.

Those of a positivist persuasion would consider empirical equivalence to be a sufficient criterion for establishing theoretical equivalence because they would argue that a theory's meaning and content is exhausted by its empirical consequences. Yet, most subscribe to the idea that empirical equivalence is a necessary but not sufficient condition for theoretical equivalence, because there are meaningful theoretical claims beyond strict empirical consequences, such as two theories differing in regards to ``what structure they attribute to the world, what sorts of entities exist in the world, or what the laws of nature are" \parencite[p.~5]{Weatherall2018-WEATEI}. This has motivated philosophers to propose further, stronger criteria for establishing theoretical equivalence that go beyond empirical consequences. These can be roughly broken down into formal notions of equivalence and interpretational equivalence. This literature is vast and we make no attempt at a fully exhaustive description of the possibilities.

\textit{Definitional equivalence} is a formal criterion developed initially by \textcite{Quine1975-QUIOEE} and \textcite{Glymour1970-GLYTRA}, and captures the idea that two theories should be inter-translatable. This means that one should be able to take all of the vocabulary of theory $T$, and translate it into the vocabulary of theory $T'$, and \emph{vice versa}, in a manner that faithfully preserves the content of each theory. Furthermore, there is generally an idea that these translations between theories should be unique and invertible. 
Other formal attempts at cashing out equivalence in something like this manner include categorical equivalence and Morita equivalence. \textit{Categorical equivalence} uses tools from category theory to address situations that seem otherwise to be problem cases for definitional equivalence, such as when transformations between models are many-to-one \parencite{Weatherall2018-WEATEI}. This is the case, for example, when multiple gauge choices in one theory correspond to one model on the other side of the transformation. \emph{Morita equivalence} attempts to weaken definitional equivalence by providing a notion of equivalence that applies to theories that are formulated using different sorts (i.e., different classes of entities) \parencite{Barrett2016-BARME-5}.

\textit{Interpretational equivalence}, in contrast with definitional equivalence, seeks to capture the notion that two theories are equivalent when they license all of the same claims about the phenomena they describe, going beyond purely empirical or formal considerations \parencite{Coffey, Teitel2021-TEIWTE}. In other words, theories $T$ and $T'$ can be understood to postulate the same ontologies and make the same claims about this shared ontology.

With these notions of equivalence on the table, we next move on to analyzing some recent discussions in the philosophical literature surrounding the issue of theoretical equivalence, and to evaluating these respective adjudications of theoretical equivalence for particular theories. The examples we will consider are (i) the equivalence of the Faraday tensor formulation and the gauge field formulation of electromagnetism \parencite{Weatherall2016-WEAANG}, and (ii) the equivalence of general relativity (GR) and the teleparallel equivalent of general relativity (TPG) \parencite{Knox2011-KNONTA} .

\section{Adjudicating Theoretical Equivalence}

\subsection{Example 1: Faraday Tensor EM and Gauge Potential EM}

We begin with the very familiar example of classical electromagnetism (EM) and consider how it has recently been presented in the philosophical literature on theory equivalence. \textcite{Weatherall2016-WEAANG} examines two different formulations of classical EM: EM$_1$, where electromagnetism is presented in terms of the electric and magnetic fields through the Faraday tensor $F_{\mu\nu}$, and EM$_2$, where electromagnetism is presented in terms of the electromagnetic gauge potential $A_{\mu}$. There is a near-universal consensus amongst both physicists and philosophers that EM$_1$ and EM$_2$ are in fact theoretically (hence empirically) equivalent.\footnote{This consensus is not quite universal. For example, once quantum mechanical considerations are taken into account, the Aharonov-Bohm effect seems to indicate that there is empirical information contained within the gauge potential that is not present in the electromagnetic fields \parencite{Aharonov:1959fk, healey}. (Conversely, \textcite{Vaidman} has argued that a full quantum mechanical treatment of source charges responsible for the electromagnetic fields can likewise account for this effect.)
%This issue does not come up in purely classical contexts though so does not affect our discussion regarding empirical equivalence.
Notwithstanding these empirical considerations, \textcite{Maudlin2018} presents the case that there are significant ontological implications (e.g.\ regarding locality, spacetime structure, etc.) that depend on whether we take the fields or potentials to be the fundamental objects of interest. These interpretational considerations could certainly lead one to conclude that these formulations are not theoretically equivalent even if they are empirically equivalent (on this, see also \parencite[Ch.~7]{North2021-NORPSA-5}).} However, due to issues that will become apparent shortly, cashing out this equivalence in terms of the criteria for theoretical equivalence that we have discussed is more subtle than one might expect. Thus, the question is not whether these formulations are truly equivalent, but rather how to state this equivalence perspicuously according to plausible criteria that physically equivalent theories should satisfy.

Weatherall proceeds by noting that these two different formulations do not meet the standard criteria for definitional equivalence as proposed by \textcite{Glymour1970-GLYTRA} due to there being non-isomorphic translations between formulations, and then arguing that one can use categorical equivalence to capture the theoretical equivalence of these formulations. However, before the argument even gets to the point at which we must decide which notion of formal equivalence is suitable for this example, it is important to emphasize that an argument for full theoretical equivalence also necessarily depends upon establishing that models derived from these different formulations really do capture \emph{all} of the same empirical content and are thus empirically equivalent as well. Within the argument, dynamical equivalence is assumed to be sufficient to demonstrate empirical equivalence, but as we shall see this seems to ignore significant empirical content that is not available in the dynamical equations of motion. Consequently, the argument fails to go through even before we get into the thornier issues surrounding the formal equivalence of these theories. To stress (and to repeat), there is an overwhelming consensus from both physicists and philosophers that EM$_1$ and EM$_2$ \textit{are} empirically equivalent. We are not challenging this, but rather challenging the philosophical criteria used in this analysis because these criteria fail to capture the empirical equivalence that these two formulations readily display within the practice of physics.

Examining Weatherall's analysis more closely, we find that he utilizes a conceptual framework that is broadly consistent with the semantic conception of scientific theories. We will have more to say on this later, but essentially the standard articulation of the semantic view holds that theories are collections of dynamically equivalent models. Weatherall takes EM$_1$ to be a theory given by models built out of the objects $\left<M, \eta_{\mu\nu}, F_{\mu\nu}, J^\mu \right>$, where $M$ is a smooth manifold, $\eta_{\mu\nu}$ is the Minkowski metric, $F_{\mu\nu}$ is the Faraday tensor, and $J^{\mu}$ is the charge density current. These models furthermore must all satisfy the dynamics encoded by Maxwell's equations
\begin{align}
\nabla_{[\rho}F_{\mu\nu]} &= 0, \\
\nabla_{\mu} F^{\mu\nu} &= J^{\nu},
\end{align}
On the other hand, EM$_2$ is a theory given by $\left<M, \eta_{\mu\nu}, A_{\mu}, J^\mu \right>$, where $A_{\mu} = (\phi, \vec{A})$ is the four-potential vector field. These models likewise satisfy Maxwell's equations in the form
\begin{equation}
\Box A^{\mu} = J^{\mu},
\end{equation}
where $\Box$ is the D'Alembertian operator.\footnote{When we refer to `Maxwell's equations' in what follows, strictly speaking we equivocate between the formulation of these equations in terms of $F_{\mu\nu}$ and the formulation in terms of $A_\mu$; that said, which version we have in mind should always be evident from context.} Weatherall's analysis quite understandably holds that these two `theories' or `formulations' of a single theory (whichever you prefer), are empirically equivalent:
\begin{description}
    \item[Empirical equivalence:] ``We stipulate that on both formulations, the empirical content of a model is exhausted by its associated Faraday tensor [that satisfies Maxwell's equations]. In this sense, the theories are empirically equivalent, since for any model of EM$_1$, there is a corresponding model of EM$_2$ with the same empirical content (for some fixed $J^a$), and vice versa" \parencite[p.~1078]{Weatherall2016-WEAANG}. In other words, EM$_1$ and EM$_2$ both share all of the same dynamical content and are thus empirically equivalent.
    %\footnote{One might be puzzled here by Weatherall's lack of mention of the Aharonov-Bohm effect on this point; we will not dwell on this issue further in this article.}
    %\todo{Does Weatherall mention the AB effect at all in his paper? If not, that seems like a major omission.}
\end{description}

Assuming that this claim of empirical equivalence goes through, one then naturally proceeds to analyze formal equivalence. As the familiar story goes, these different formulations are very closely related. Given the Faraday tensor $F_{\mu\nu}$ that satisfies Maxwell's equations, there is always a vector field $A_{\mu}$ that also satisfies Maxwell's equations and satisfies
\begin{equation}\label{eq:FA}
F_{\mu\nu} = \nabla_{[\mu}A_{\nu]}.
\end{equation}
Similarly, given a vector field $A_{\mu}$ that satisfies Maxwell's equations, there is always a corresponding tensor $F_{\mu\nu}$ that satisfies Maxwell's equations and can be defined via \eqref{eq:FA} (all of these facts follow from elementary properties of differential forms). As Weatherall notes, however, one cannot find an isomorphism between the spaces of models of these two formulations of classical electromagnetism. Starting with the EM$_2$ formulation and given a vector potential $A_{\mu}$, one can uniquely define a Faraday tensor $F_{\mu\nu}$ in EM$_1$. Conversely, going in the other direction and given a Faraday tensor $F_{\mu\nu}$ in EM$_1$, one cannot uniquely determine a model in EM$_2$ due to the gauge freedom present in the four-potential $A_{\mu}$. That is, $F_{\mu\nu}$ is compatible with infinitely many different $A_{\mu}$ because \eqref{eq:FA} will hold for any $A_{\mu}$ such that 
\begin{equation}
A'_{\mu} = A_{\mu} + G_{\mu}, \qquad \nabla_{[\mu}G_{\nu]} = 0
\end{equation}
(i.e.\ if $G_{\mu}$ is a closed one-form). Given that a straightforward application of definitional equivalence is blocked, Weatherall motivates abandoning Glymour's criterion for definitional equivalence (i.e., for every model in $T$, there is an isomorphic translation to a model in $T'$ that preserves all of the same empirical content) in favor of demonstrating an equivalence between categories of models that preserve empirical content. Here, we now understand the models of EM$_2$ to be $\left<M, \eta_{\mu\nu}, [A_{\mu}], J^\mu \right>$, where $[A_{\mu}]$ is understood as an ``equivalence class of physically equivalent vector potentials" that correspond to the same $F_{\mu\nu}$ \parencite[p.~1079]{Weatherall2016-WEAANG}. Note that this adjustment depends on the argument that EM$_1$ and EM$_2$ are actually empirically equivalent, and this equivalence class of vector potentials contains identical empirical information as its counterpart in the corresponding field formulation.

\begin{description}
    \item[Categorical equivalence:] Categorical equivalence is stated in terms of categories of models that preserve empirical content. Thus, according to Weatherall's construction, we can translate between models of EM$_1$ and EM$_2$ and their respective vocabularies in a manner that faithfully preserves empirical content, provided that EM$_2$ is redefined such that $[A_{\mu}]$ is an equivalence class of vector potentials that lead to the same $F_{\mu\nu}$. Then, we have $\left<M, \eta_{\mu\nu}, F_{\mu\nu}, J^\mu \right> \Longleftrightarrow \left<M, \eta_{\mu\nu}, [A_{\mu}], J^\mu \right>$, and this further notion of formal equivalence is then used to argue that both formulations are theoretically equivalent \parencite[p.~1083]{Weatherall2016-WEAANG}.
\end{description}

While it is certainly correct that all models in both formulations possess the same dynamical content, as we shall soon see, this does not mean that they necessarily share \textit{all} of the same empirical content. Consider a simple environment-subsystem decomposition that includes a basic Faraday cage, described by a finite spatial subsystem region with a surface boundary $\partial M$. The Faraday cage is a perfect electrical conductor, meaning that it effectively shields the subsystem from electromagnetic fields in the environment and any electric charge is accumulated on the boundary in the form of a surface charge $\sigma$. When decomposing the environment and subsystem, boundary conditions delineate the subsystem from the environment. There are two relevant boundary conditions in this example: (1) $E_{\parallel} = 0$, meaning that the electric field vanishes on the boundary in the direction parallel to the surface and (2) $E_{\perp} = 4\pi\sigma$, meaning that the electric field is proportional to the surface charge in the direction perpendicular to the surface.\footnote{See \textcite{zangwill_2012} or \textcite{griffiths_2017} for detailed discussions of perfect conductors, boundary conditions, and related topics.} Let us now consider EM$_1$ and EM$_2$ models of the Faraday cage subsystem.

Beginning with the Faraday tensor formulation EM$_1$, this construction in terms of the electric and magnetic fields will lead to the conclusion that the Faraday tensor describing the subsystem is \textit{always} zero. This is simply a consequence of the fact that regardless of what the external electric and magnetic fields are, the conducting boundary will always arrange the surface charge $\sigma$ to cancel the effect of the external fields. Thus, $F_{\mu\nu} = 0$ inside the cage regardless of facts about the external fields and surface charge. By contrast, the gauge field formulation EM$_2$ shows that the gauge potentials describing the subsystem will instead be constant, which is a result of the simple fact that the Faraday tensor is zero and any points lying inside the conductor must then lie at the same potential. While so far this is what we expect of potentials that lead to $F_{\mu\nu} = 0$, it is also true that specifying the scalar electric potential $\phi$ on the boundary (and thus the potential for the entire interior) \textit{uniquely} specifies the surface charge $\sigma$ on the boundary \parencite[p.~200]{zangwill_2012}.\footnote{\textcite{MurgueitioRamirezForthcoming-MURAGS-3} have also emphasized this point in their paper concerning the direct empirical significance of gauge symmetries.} Furthermore, in general one can fully construct a solution for $\phi$ for both the subsystem and exterior in terms of the surface charge $\sigma$.\footnote{For example, consider one of the simplest perfect conductors: a hollow spherical shell of radius $R$ with a surface charge $q = 4 \pi R^2\sigma$. The electric potential is given by \parencite[87]{griffiths_2017}: 
\begin{equation}
V(r)= \begin{cases}\frac{1}{4 \pi \epsilon_0} \frac{q}{r} & (r \geq R), \\ \frac{1}{4 \pi \epsilon_0} \frac{q}{R} & (r \leq R). \end{cases}
\end{equation}
Here we see the interior potential ($r \leq R $) is a constant function of the surface boundary charge.}
%, which comes back to the fact that shifts in the potential shift the charge at the boundary surface.\footnote{This happens in three steps, following \textcite{MurgueitioRamirezForthcoming-MURAGS-3}. (1) One introduces a gauge transformation $A \rightarrow A' + d \chi (x,t)$, where $\chi$ is the gauge parameter. (2) One fixes the gauge parameter by choosing the Coulomb gauge by $\nabla \cdot A' = \nabla \cdot (A + \nabla \chi (x,t) ) = 0$. (3) This then leads to Poisson's equation for the scalar electric potential $\phi (x)$. They note that this procedure reveals interesting features of the gauge parameter $\chi (x,t)$. That is, $\chi (x,t)$ is a field-dependent parameter that depends on the four-potential $A$ and in gauge-fixing this parameter, one shifts the electric scalar potential $\phi(x) \mapsto \phi'(x)=\phi(x)-\partial_{t} \chi(x, t)$ both in the interior and boundary of the subsystem. To make this more explicit, $\phi'$ satisfies the Poisson equation, where at the boundary $\nabla^2 \phi' = \sigma$. We can then recognize a relationship between $\chi$ and $\sigma$ through $\nabla^2 (\phi(x)-\partial_{t} \chi(x, t)) = \sigma$. As specifying the potential $\phi$ uniquely specifies the surface charge $\sigma$, shifts in the gauge parameter $\chi$ also induce shifts in the surface charge $\sigma$.}

How do these considerations influence our verdict on the empirical equivalence of these two formulations? EM$_1$ treats the Faraday tensor as the  fundamental object of interest. The same Faraday tensor $F_{\mu\nu}$ within the isolated subsystem could potentially correspond to two empirically distinct surface charges $\sigma_1$ and $\sigma_2$ (in fact, it corresponds to infinitely many distinct surface charges!). However, EM$_2$ treats gauge potentials as the fundamental objects of interest. Once we construct the gauge potential $\phi$, it will always distinguish between $\sigma_1$ and $\sigma_2$ because specifying the potential uniquely specifies the surface charge. To be completely explicit, let us adopt Weatherall's initial characterization of EM$_1$ and EM$_2$ as models given by $\left<M, \eta_{\mu\nu}, F_{\mu\nu}, J^\mu \right>$ and $\left<M, \eta_{\mu\nu}, A_{\mu}, J^\mu \right>$, respectively.\footnote{Here we are concerned with empirical equivalence which is independent of the categorical equivalence issue that Weatherall addresses in his later characterization of EM$_2$.} Furthermore, let us say that we are interested in an empirical description of a Faraday cage with a surface charge $\sigma_1$. On Weatherall's characterization, EM$_1$ corresponds to $\left<M, \eta_{\mu\nu}, 0, J^\mu \right>$ and EM$_2$ corresponds to $\left<M, \eta_{\mu\nu}, \phi_s(\sigma_1) , J^\mu  \right>$, where $\phi_s$ is the scalar potential for the subsystem. This EM$_1$ description could correspond to infinitely many subsystems all with different surface charges because they will all lead to $F_{\mu\nu}=0$, whereas the EM$_2$ description uniquely describes the subsystem with the particular surface charge we are considering here. In other words, the $U(1)$ gauge orbit of models within EM$_2$ each make a specific assertion about the surface charge because the potential $\phi$ fixes the surface charge uniquely. The dynamically equivalent counterparts within EM$_1$ are completely silent on the matter. Consequently, the model of the subsystem in EM$_2$ has the information necessary to model empirical facts about boundary phenomena and the external environment, empirical information that the model of the subsystem in EM$_1$ simply does not have when we hold that the empirical content of the theory is given exclusively in terms of those particular mathematical objects and their dynamics. On this reading, these descriptions of the subsystem are not empirically equivalent because they do not carry the same empirical information about the target system, nor can the same empirical consequences be deduced from them.\footnote{One could say this: the additional structure present in the models of EM$_2$ decreases representational freedom with respect to boundary phenomena, thereby breaking empirical equivalence with EM$_1$. cf.~\textcite[p.~272]{BradleyWeatherall}.}

It is important to emphasize here that we are not arguing against the actual empirical equivalence of the Faraday tensor and gauge field formulations of electromagnetism. Essentially, a physicist can deduce the same empirical claims about such a subsystem from both formulations as using these formulations in practice involves specifying further items (like the boundary conditions and their relationship to the surface charges) that are necessary to build the electromagnetic fields and potentials relevant to describing the system. Rather, we are arguing that the philosophical criterion for evaluating empirical equivalence in terms of dynamics alone is insufficient to account for the equivalence of EM$_1$ and EM$_2$. Indeed, this view leads to the conclusion that EM$_1$ and EM$_2$ (when stated as consisting of models $\left<M, \eta_{\mu\nu}, F_{\mu\nu}, J^\mu \right>$ and $\left<M, \eta_{\mu\nu}, A_{\mu}, J^\mu \right>$ respectively) are \emph{not} equivalent because there is significantly more empirical information contained within a model specified as $\left<M, \eta_{\mu\nu}, A_{\mu}, J^\mu \right>$. Here, the attempt to demonstrate theoretical equivalence gets tripped up, not in the nuanced technicalities surrounding formal notions of equivalence, but rather in the more mundane issue of empirical equivalence. This suggests that dynamical equivalence is not sufficient for empirical equivalence and that there is further empirical information that must be added to secure a verdict of equivalence. As we shall argue in \S\ref{BPM}, there is a relatively straightforward way of modifying our philosophical view of theory structure which can bring the empirical claims from these formulations into alignment (again from the perspective of the \textit{philosophical} criteria we are employing) and restore the near-universal intuition that these formulations are in fact equivalent \emph{tout court}. 

%In this simple example, it is easy to see that we can understand both the Faraday tensor and gauge field formulations as empirically equivalent if we think of the boundary conditions listed above as empirical information that is relevant to such a determination. The point of this exercise is to demonstrate that this particular empirical information is not available in the dynamics of the theory. Here, the attempt to demonstrate theoretical equivalence gets tripped up, not in the nuanced technicalities surrounding formal notions of equivalent, but rather in the more mundane issue of empirical equivalence. We can indeed argue that these theories are empirically equivalent and restore our very plausible intuitions that they are equivalent full stop, but in some cases we must add further empirical information that goes beyond dynamics in order to secure this verdict. 

\subsection{Example 2: GR and TPG}
The example above is not the only instance from the recent philosophical literature where there has been a proclaimed equivalence between two theories that relies on understanding dynamical equivalence as being sufficient for complete empirical equivalence. This has also come up in the context of general relativity (GR) and the teleparallel equivalent of general relativity (TPG) \parencite{Knox2011-KNONTA}.

Both GR and TPG are theories of gravitation, but they differ in a number of ways. The most obvious is that rather than using the curved, symmetric Levi-Civita connection $\Gamma^{\rho}_{\mu\nu}$,  TPG uses the Weitzenb\"{o}ck connection $\dot{\Gamma}^{\rho}_{\mu\nu}$, which has non-vanishing torsion and vanishing curvature.\footnote{It should be noted that this does not in any way exhaust the possibilities for the geometric properties of connections. Indeed, there are also connections which utilize the geometric concept of non-metricity, rather than torsion or curvature. While we will not address this here, see \textcite{BeltranJimenez:2019esp, Wolf:2023rad} for further physics discussion of the symmetric teleparallel equivalent of general relativity (STG). This theory, like TPG, attempts to reproduce the predictions of GR while using a connection with vanishing torsion and curvature, but non-vanishing non-metricity. Curvature, torsion, and non-metricity together form a so-called `geometrical trinity'. See \textcite{Wolf:2023xrv} for further philosophical discussion.} That is, rather than expressing gravity as a manifestation of spacetime curvature as GR does, TPG holds that gravity is a manifestation of spacetime torsion. TPG views gravity as a force because torsion directs bodies experiencing gravitation away from geodesics, as opposed to the situation in GR, whereby bodies experiencing gravitation follow the geodesics resulting from spacetime curvature. Furthermore, TPG is usually formulated in terms of tetrads $e^a_\mu$, rather than a metric $g_{\mu\nu}$. Tetrads, or frame fields, are sets of four linearly independent fields $e^a = e^a_\mu dx^\mu$ that at each point $p$ of a differentiable manifold $M$ specify a basis for the tangent space $T_p M$.\footnote{What we have in fact written here are cotetrad fields $e^a$, which are those 1-forms such that $e^a_\mu e_b^\mu = \delta^a_b$; we do so since this will simplify the presentation in what follows.} TPG uses frame fields $h^a_\mu = e^a_\mu + B^a_\mu$ that are constructed to be invariant under local translations $x^a \mapsto x^a + \epsilon^a$, where $B^a_\mu$ is the translation gauge potential. This gauge potential transforms as $\delta B^a_\mu = -\partial_\mu \epsilon^a$ so as to make the frame field invariant under such local translations. It is for this reason that TPG is often declared to be a gauge theory of the translation group \parencite{Aldrovandi:2013wha}.\footnote{For some critical discussion of this claim, see \textcite{WallaceFields}.}

GR and TPG are seemingly very distinct theories, constructed using different mathematical structures---but \textcite{Knox2011-KNONTA} has argued that GR and TPG should in fact be understood as being equivalent to one another. She argues for this conclusion based upon: (i) the establishment of dynamical equivalence (and thus, for her argument, empirical equivalence) and definitional equivalence between the two theories, and (ii) an interpretation of TPG that holds that both TPG and GR actually postulate the same underlying spacetime structure despite the surface level appearances, which appears to be motivated by her advocacy of spacetime functionalism. As before, theoretical equivalence is taken to be a combination of demonstrating empirical equivalence, along with some stronger notions of equivalence that demonstrate clear formal relations between the theories or resolve interpretive issues such that we can understand both theories as making the same claims about the target phenomena. While the spacetime functionalist component of her argument certainly brings up a host of interesting issues, this is not the place to fully adjudicate the interpretational issues she raises regarding TPG and GR. However, we would like to focus specifically on the discussion of empirical equivalence between the theories.

The claim that TPG and GR are empirically equivalent is motivated by appealing to actions used in each theory,
%\todo{Ruward suggested we should use the torsion scalar here---what do you think?}
\begin{align}
    S_{TPG} &= \frac{1}{16 \pi G} \int d^4x \sqrt{h} T, & S_{GR} &= \frac{1}{16 \pi G} \int d^4x \sqrt{g} R,
\end{align}
% \begin{align}
%     S_{TPG} &= \frac{1}{16 \pi G} \int d^4x \sqrt{e} \mathcal{S}_\rho^{\mu\nu} T^\rho_{\mu\nu}, & S_{GR} &= \frac{1}{16 \pi G} \int d^4x \sqrt{g} R,
% \end{align}
where ${h}$ is the determinant of the tetrad, $T$ is the torsion scalar defined as
\begin{equation}
T := \mathcal{S}_\rho^{\mu\nu} T^\rho_{\mu\nu},
\end{equation}
$\mathcal{S}_\rho^{\mu\nu}$ is the so-called superpotential tensor, $T^\rho_{\mu\nu}$ is the torsion tensor, ${g}$ is the determinant of the metric, and $R$ is the Ricci scalar. The superpotential tensor is built out of the torsion tensor and the so-called contorsion tensor, which is defined as
\begin{equation}
K^{\rho}_{\mu\nu} := \Gamma^{\rho}_{\mu\nu} - \dot{\Gamma}^{\rho}_{\mu\nu},
\end{equation}
where we see that it is simply the difference between the Weitzenbock connection, $ \dot{\Gamma}^{\rho}_{\mu\nu}$, and the Levi-Civita connection, $\Gamma^{\rho}_{\mu\nu}$. This is significant because this allows one to translate between the mathematical structures of the teleparallel theory and those of general relativity. One can use this to re-write the TPG action in the language of GR as\footnote{See e.g.~\textcite[Ch.~9]{Aldrovandi:2013wha} for a fully explicit derivation.}
% \begin{equation}
%     S_{TPG} = \frac{1}{16 \pi G} \int d^4x \sqrt{g} R + \frac{1}{8 \pi G} \oint_{\partial M} d^{3} \Omega \sqrt{h} n_{\mu} T_{\alpha}^{\alpha \mu}.
% \end{equation}
\begin{equation}\label{EHTPG1}
    S_{TPG} = \frac{1}{16 \pi G} \int d^4x \sqrt{g} R + \frac{1}{8 \pi G} \int d^{4}x \sqrt{g} \nabla_\mu T_{\alpha}^{\alpha \mu}.
\end{equation}
This shows that the TPG action is identical to the Einstein-Hilbert action of GR plus a total divergence term, which ensures that these actions both lead to the same dynamical equations of motion. On the basis of these observations, Knox makes three arguments regarding the equivalence of GR and TPG:
%\todo{Can we give page citations for each of these claims we're imputing to Knox?} \todoWill{Done}
\begin{description}
    \item[Empirical equivalence:] The equivalence of the actions up to a total divergence term, which indicates that they both share equivalent equations of motion, guarantees the empirical equivalence of TPG and GR \parencite[p.~272]{Knox2011-KNONTA}.
    \item[Definitional equivalence:] The relationship between the Levi-Civita connection and the Weizenb\"{o}ck connection allows us to directly translate between GR and TPG and \emph{vice versa}. While definitional equivalence is not explicitly mentioned in her argument, this is a clear appeal to a similar notion of equivalence. Anything we express in the language of GR can be equivalently expressed in the language of TPG and \emph{vice versa} in a way that preserves the content of each theory. For example, we have already seen how one moves between different connection coefficients and translates between spacetime curvature and torsion, but one can similarly translate between the frame fields of TPG and the metric of GR as $g_{\mu\nu} = \eta_{ab}h^a_\mu h^b_\nu$, where $\eta_{ab}$ is the Minkowski metric \parencite[p.~272]{Knox2011-KNONTA}.
    \item[Interpretational equivalence:] TPG and GR both encode the same spacetime structure, upon adopting spacetime functionalism (which, for Knox, is the view that spacetime structure is whatever identifies a class of local inertial frames---for critical discussion of this view, see e.g.~\parencite{MR}), and thus can be understood as licensing the same claims about the phenomena they describe \parencite[p.~273]{Knox2011-KNONTA}.
\end{description}

The argument that the actions are empirically equivalent hinges on the ability to throw away the total divergence term present in \eqref{EHTPG1}. Once this term is discarded, the actions are equivalent full stop and the argument for definitional equivalence goes through as well because these terms can be safely ignored when making these kinds of translations between TPG and GR. But why can this total divergence term simply be thrown away? 

When discussing a particular theory whose content is captured by an action $S$, typically one takes the empirical content of that theory to be derived from a variational principle.\footnote{This isn't to say that variational principles \emph{exhaust} the empirical content of theories---for example, conserved quantities can be derived via Noether's theorems. These points will not matter for our purposes here.} The `principle of least action' is a variational principle which holds that the variation of the action is held fixed when the equations of motion---i.e.,~the dynamics---of the system are satisfied. Consider the simple textbook example of a free massive particle in motion where our variables are position $q (t)$ and velocity $\dot{q}(t)$ and the action is given by  $S = \int_{t_i}^{t_f} L[q, \dot{q}, t] dt$:
\begin{align}
    \delta S=\int_{t_{i}}^{t_{f}}\left[\frac{\partial L}{\partial q}-\frac{d}{d t}\left(\frac{\partial L}{\partial \dot{q}}\right)\right] \delta q d t+\frac{\partial L}{\partial \dot{q}}\left(t_{f}\right) \delta q\left(t_{f}\right)-\frac{\partial L}{\partial \dot{q}}\left(t_{i}\right) \delta q\left(t_{i}\right)=0.
\end{align}

Here we find the familiar Euler-Lagrange equations of motion in the first term. However, we also have two further terms which are the result of a total divergence that appears after the integration by parts necessary to write the Euler-Lagrange equations in their standard form. In this case we are simply concerned with the motion of a particle between two fixed end points, $\delta q (t_i)$ an $\delta q (t_f)$. These remaining terms thus automatically go to zero, leaving just the dynamics of our system captured in the first term. These total divergence terms do not affect the underlying dynamics of the system; furthermore, 
%. That is, one could introduce any number of these terms without impacting the dynamical content of the theory in question. This is because 
it is important to emphasize that any terms like this \textit{must} vanish for there to be a well-defined variational principle at all, as a proper functional derivative could not be defined otherwise.
%If they do not vanish, the variational principle is not well-defined and such a procedure does not capture any empirical content because the functional $S$ itself is not even differentiable. \todo{I agree, but can we perhaps be slightly more explicit about the reasons underlying this final sentence?} \todoWill{The literature and stack exchange makes this statement but i dont quite get why because no one actually explains it.}

Given that we typically throw away total divergence terms because we know that they have to vanish anyway, our work, apparently, is done. The TPG action encodes the same dynamics as the GR action, so the equations of motion will be the same and we are left to choose the language in which to express them: the force equations of TPG or the geodesic equations of GR. That is,
\begin{equation}
\delta S_{GR} = \delta S_{TPG} = \frac{1}{16 \pi G}\int d^4x\sqrt{-g}G_{\mu\nu}\delta g^{\mu\nu},
\end{equation}
where $G_{\mu\nu}$ contains the dynamical equations of motion. Thus, ``the equivalence of the Lagrangians is enough to establish empirical equivalence" \parencite[p.~272]{Knox2011-KNONTA}. The question of the theoretical equivalence between TPG and GR then hinges only upon the interpretive questions.

When doing GR, we often consider manifolds without boundary. This guarantees that the total divergence term in \eqref{EHTPG1} is zero because Stokes' theorem allows us to convert a total divergence term into a boundary term. In the event that there is no boundary, this term vanishes automatically. For example, this is exactly what is done in using GR to model cosmological solutions as we are attempting to model the entire universe and its contents filling an infinite space. What if we wanted to model some isolated subsystem instead? Consider an isolated subsystem $\mathcal{S}$ that is being modelled with respect to an external environment $\mathcal{E}$. For example, we might be interested in describing the mass-energy content of a region of spacetime, such as the mass-energy content contained within a black hole, as defined by an external observer who is sufficiently far away so that they do not interact with any of the relevant gravitational or material fields. In this event, it is not appropriate to consider manifolds without boundary. Rather, the manifold $M$ must have a boundary $\partial M$ along with appropriate boundary conditions to properly describe a subsystem $\mathcal{S}$ isolated from its environment $\mathcal{E}$. Total divergence terms such as the one we have considered then cannot be automatically discarded and generally will not vanish.

% As mentioned before, in the scenario that the manifold $M$ has no boundary, this boundary term simply vanishes, which yields,
% \begin{equation}
%      \delta S_{\mathrm{GR}} =\frac{1}{16 \pi G} \int_{M} d^{4} x \sqrt{-g} G_{\mu \nu} \delta g^{\mu \nu}.
% \end{equation}
% This of course squares with everything we have said as the variation of the action is zero when the dynamics given by $G_{\mu\nu}$ is satisfied and there is a well-defined variational principle.

When considering the Einstein-Hilbert action in the presence of the boundary $\partial M$, such residual total divergence terms are indeed present and we must find appropriate boundary conditions to render this a well-defined variation.\footnote{Recalling the Einstein-Hilbert action of GR, considering a manifold with a boundary $\partial M$, and varying the action yields
% \begin{equation}
%     S_{\mathrm{GR}}=\frac{1}{16 \pi G} \int d^{4} x \sqrt{-g} R,
% \end{equation} 
\begin{equation}\label{varyEH}
    \delta S_{\mathrm{GR}} =\frac{1}{16 \pi G} \int_{M} d^{4} x \sqrt{-g} G_{\mu \nu} \delta g^{\mu \nu} +\frac{1}{16 \pi G} \oint_{\partial M} d^{3} \Omega \sqrt{h} \left(g^{\sigma \nu} \delta \Gamma_{\nu \sigma}^{\rho}-g^{\sigma \rho} \delta \Gamma_{\mu \sigma}^{\mu}\right),
\end{equation}
where $G_{\mu\nu}$ is the Einstein tensor, $g_{\mu\nu}$ is the metric tensor, $h_{\mu\nu}$ is the induced metric on the boundary, and $\Gamma^{\mu}_{\nu\lambda}$ represents the connection coefficients of the Levi-Civita connection \textcite[ch.~20.2]{Blau}. The first term yields via variation the Einstein field equations and vanishes when the dynamics of the theory are satisfied. The remaining term comes from a total divergence that has been converted to a boundary term via Stokes' theorem.} Here, it is natural to consider Dirichlet boundary conditions, $\left.\delta g_{\mu \nu}\right|_{\partial M}=0$, as these boundary conditions are often used in the context of asymptotically flat spacetimes. These are spacetimes that approach flatness $g_{\mu\nu} \rightarrow \eta_{\mu\nu}$ at null-infinity and are particularly significant for a number of reasons. Here is \textcite{Penrose1982SOMEUP} on the issue:
\begin{quote}
    Asymptotically flat spacetimes are interesting, not because they are thought to be realistic models for the entire universe, but because they describe the gravitational fields of isolated systems, and because it is only with asymptotic flatness that general relativity begins to relate in a clear way to many of the important aspects of the rest of physics, such as energy, momentum, radiation, etc.
\end{quote}
That is, in the asymptotic regime we can clearly define critical, empirically relevant concepts such as mass, energy, and momentum, and relate them to these concepts as they are understood in other realms of physics. (In brief: in the asymptotic regime, one has Killing fields, with which one can associate conserved quantities in a well-understood way: see e.g.~\textcite{DeHaro:2021gdv}.)

%(sentence or two about the intuition here)\todo{What do you have in mind here? Mention of symmetries, Noether's theorems, etc.?}\todoWill{Maybe just another sentence on \textit{why} asymptotic flatness makes a concept like energy well-defined as it is in say special relativity. I kinda get it but not entirely. Do you think it is also worth talking about Noether's theorems here? And bringing Noether charges into alignment with Hamiltonian charges as Nic's paper shows?}

Upon imposing Dirichlet boundary conditions $\left.\delta g_{\mu \nu}\right|_{\partial M}=0$, we find that there is a problem. There are multiple boundary terms and it is only the term that depends on the tangential derivatives of the metric that vanishes, while another term that depends on the normal derivatives survives.\footnote{It is helpful to rewrite the boundary term from the GR action eq.~(\ref{varyEH}) as:
\begin{equation}\label{EHboundary}
    \frac{1}{16 \pi G} \oint_{\partial M} d^{3} \Omega \sqrt{h} \left(N^{\rho} h^{\mu \nu} \nabla_{\mu} \delta g_{\rho \nu}-N^{\mu} h^{\rho \nu} \nabla_{\mu} \delta g_{\rho \nu}\right),
\end{equation}
where $N$ is the vector normal to the boundary \textcite[ch.~20.5]{Blau}. The first term depends on tangential derivatives of the metric and its variation, whereas the second term depends on normal derivatives of the metric and its variation. Dirichlet boundary conditions kill only the first term, while leaving the second term intact.} This is because Dirichlet boundary conditions fix only the values of the metric of the boundary, but this does not necessarily require that the derivatives of the metric vanish. In other words, the variation of this action does not yield a well-defined variation and cannot be used to represent or model isolated subsystems of the type that Penrose refers to in his description of asymptotically flat spacetimes. This is closely related to what  \textcite{Belot2018-GORFME} observes when he notes that two isomorphic solutions in GR do not always represent the same physical possibilities. For example, he notes that while cosmological solutions and asymptotically flat solutions are isomorphic dynamically, they (obviously) do not represent the same physical possibilities. The boundary conditions imposed for each solution are physically relevant facts! This discussion of the variational problem in GR reveals that the Einstein-Hilbert action only has the resources to represent one of the two physical possibilities we have mentioned (cosmological solutions), and that we need to look elsewhere to represent asymptotically flat solutions. We see that even within GR, dynamically equivalent solutions do not necessarily represent the same physical possibility. Thus, merely demonstrating the dynamical equivalence between a GR action and a TPG action likewise would not necessarily indicate that the two theories are physically equivalent.

This indicates the importance of boundary conditions in specifying the content of our theory and the scope of the empirical scenarios and target systems that our models and theories can represent. Let us now compare the analogous scenario in TPG to see how the teleparallel theory fares in describing isolated subsystems with aysmptotic characteristics.

Amazingly, upon varying the TPG action and imposing Dirichlet boundary conditions, we find that the TPG action indeed does have a well-defined variation \parencite{Oshita:2017nhn}. The variation of the additional boundary term that distinguishes the TPG and Einstein-Hilbert actions ensures that the total variation is well-defined for asymptotic spacetimes because the additional terms perfectly cancel out the previously problematic terms.\footnote{Recall that TPG differs from the GR action by a total divergence term. Therefore, we can apply Stokes' theorem to the divergence term in \eqref{EHTPG1}, and add this to the result above. Converting the boundary term in \eqref{varyEH} to the language of TPG frame fields, adding the additional TPG boundary term, and imposing Dirichlet boundary conditions $\left.\delta g_{\mu \nu}\right|_{\partial M} = 0, \left.\delta e^a_{\mu}\right|_{\partial M} = 0$ yields the following \parencite{Oshita:2017nhn}:
\begin{align*}
    \delta S_{\mathrm{TPG}} &=\frac{1}{16 \pi G} \int_{M} d^{4} x \sqrt{-g} G_{\mu \nu} \delta g^{\mu \nu} +\frac{\epsilon}{8 \pi G} \oint_{\partial M} d^{3} \Omega \sqrt{h} n^{\mu}\left[e_{A}^{\alpha} \partial_{\alpha} \delta e_{\mu}^{A}-e_{A}^{\alpha} \partial_{\mu} \delta e_{\alpha}^{A}\right] \\
    & \qquad + \frac{\epsilon}{8 \pi G} \oint_{\partial M} d^{3} \Omega \sqrt{h} n^{\mu}\left[e_{A}^{\alpha} \partial_{\mu} \delta e_{\alpha}^{A}-e_{A}^{\alpha} \partial_{\alpha} \delta e_{\mu}^{A}\right] \\
    &= 0.
\end{align*}} The reason for this can be traced to the fact that the TPG action contains only first derivatives of the frame fields, whereas the Einstein-Hilbert formulation contains second derivatives of the metric. The additional boundary term effectively removes the second derivatives of the metric that fail to vanish when working with the Einstien-Hilbert action.

This TPG action functions perfectly well for describing such isolated subsystems. As this specific argument for theoretical equivalence is presently formulated (relying on the dynamical equivalence of two different actions), TPG and GR are \emph{not} empirically equivalent---and so, \emph{per} the above, should not be regarded as being equivalent, full stop. Under this articulation, these theories do not even have the resources to model all of the same target systems, much less discuss whether one can compare the empirical consequences derived from them for said target systems. As before, the account of empirical equivalence gets tripped up when considering isolated subsystems and it seems like merely showing the dynamical equivalence of models derived from particular actions used in the respective theories is not enough to ensure that they can support the same empirical claims. For anyone who may understandably be perturbed by the thought that GR cannot describe such systems: do not worry. This will be addressed in \S\ref{BPM}, where we will argue that we can make an argument for the empirical equivalence of GR and TPG. However, as is also the case with the example of electromagnetism, this will require modifying our philosophical criteria concerning theory structure and considering empirical information that goes beyond mere dynamical equivalence between models.

\section{Views on Theory Structure}

\noindent What is happening here? We have two fairly prominent examples of arguments for the theoretical equivalence of the respective theories considered in these examples. One of these examples (TPG and GR) is more contentious given the extent of the interpretive arguments that need to be made to secure interpretational equivalence, but the other (Faraday tensor and vector potential formulations of EM) is fairly uncontroversial. Yet, as articulated, these arguments for theoretical equivalence cannot even support claims of empirical equivalence for these respective theories. Something has clearly gone wrong! 

Perhaps it is the way in which the theories have been stated that has disrupted these claims of empirical equivalence. After all, in making an adjudication of theoretical equivalence, it is certainly important to specify correctly the empirical content contained by a theory. Views on the structure of scientific theories can be roughly broken down into three camps: the `syntactic', `semantic', and `pragmatic' views. The syntactic view seeks to axiomatize a theory in terms of abstract mathematical sentences. The semantic view casts a theory in terms of models and the kinds of mathematical objects that comprise these models. While the syntactic view was initially dominant as it emerged first as an outgrowth from logical empiricism, van Fraassen has prominently advocated for the semantic view by arguing that the semantic view, with its focus on models, can often more simply demonstrate the logical claims of a theory than a set of axioms.\footnote{Indeed, van Fraassen acknowledges that one can often derive the same logical claims concerning the statements a theory makes about the world from both approaches, with the caveat that these claims are more clearly and simply expressed on the semantic view. \textcite{Lutz2017-LUTWWT} has taken this further and argued that both syntactic and semantic views are actually far more closely related than has been supposed in the literature and the debate surrounding which approach is preferable is largely illusory.} Furthermore, he argues that the semantic view is a far more comprehensive and useful tool because it avoids the restrictions inherent to describing a theory in a particular axiomatic language, and allows us to conceptualize the objects and classes of structures that comprise a model in terms of a variety of valid, non-unique descriptions \parencite[p.~43--4]{VanFraassenBas1980-VANTSI}.
Finally, the pragmatic view is a more recent perspective that emphasizes representational aims, model pluralism, scientific practice, and other non-formal characteristics \parencite{Cartwright1983-CARHTL, Hacking1983-HACRAI-7, Kitcher1993-KITTAO-2, sep-structure-scientific-theories}. 

In this article, we will focus on viewing these adjudications of theoretical equivalence through the lenses of both the semantic and pragmatic views. While neither Weatherall nor Knox explicitly advocates a particular view of theory structure, both authors' focus on dynamics and models reflects at the very least a straightforward consistency with fairly standard articulations of the semantic view, making this a natural place to start. Regarding the syntactic view, it should be noted that nothing automatically precludes a discussion in syntactic-friendly terms; however, these authors do not engage with this approach in any obvious way, so likewise we will not do so here. And finally, we will engage with the pragmatic view, as its focus on model pluralism and scientific practice is particularly relevant for the questions at hand and arguably can shed some light on these adjudications of equivalence.

\subsection{The Semantic View}

The semantic view of theories holds that a theory is individuated via classes of models. One modern way of expressing the semantic view is to say that a theory $\mathcal{T}$ has a set of `kinematically possible models' $\mathcal{K}$ (KPMs), defined by tuples of the form $\left<O_i, ... O_n\right>$, where these $O_i$ are mathematical objects, e.g.~tensor fields on a differentiable manifold. Furthermore, these objects come with a set of particular dynamical equations that define the relationships and interactions between the $O_i$. KPMs that satisfy these dynamical equations form a subspace $\mathcal{D} \subset \mathcal{K}$ of KPMs known as the `dynamically possible models' (DPMs). 
In other words, ``the KPMs can be thought of as representing the range of metaphysical possibilities consistent with the theory’s basic ontological assumptions. The DPMs represent a narrower set of physical possibilities" \parencite[p.~532]{Pooley2013-POOSAR}. This dynamical content is then understood to capture the empirical content of the models that comprise the theory, via what van Fraassen calls the `empirical substructures' of each of these models \parencite[p.~45]{VanFraassenBas1980-VANTSI}.

It is clear that Weatherall draws from this framework in his analysis. For example, his descriptions of EM$_1$ and EM$_2$ as theories with associated respective classes of models $\left<M, \eta_{\mu\nu}, F_{\mu\nu}, J^\mu \right>$ and $\left<M, \eta_{\mu\nu}, A_{\mu}, J^\mu \right>$ identifies the relevant KPMs, where his specification that these models obey Maxwell's equations identifies the particular DPMs that correspond to the theories in question.

While the utilization of the standard semantic view is not as obvious in Knox,\footnote{Knox does not explicitly construct KPMs and DPMs of GR and TPG respectively; however, she does discuss KPMs and DPMs of Newtonian gravity and Newton-Cartan theory in fairly straightforward semantic terms \parencite[\S1]{Knox2011-KNONTA}.} it is clear that something like this is being supposed in her identifying the theory of GR with the empirical content contained within the Einstein-Hilbert action. Recall that in her argument it is the local equivalence of the two actions that cements the case for empirical equivalence, which really is just the statement that both theories share the same dynamical content when the actions are varied per standard variational principles. In identifying the Einstein-Hilbert action as capturing GR's content and adjudicating the empirical equivalence of GR and TPG based on the dynamical equivalence of these actions, there is a naturally consistency with the standard semantic expression of GR in the philosophical literature.

In more detail: in the above-introduced model-based language \parencite{Pooley2013-POOSAR, Pooley2015-POOBID}, GR is usually given by KPMs of the form $\left<M, g_{\mu\nu}, \Phi\right>$, where (again) $M$ is a smooth, four dimensional differentiable manifold, $g_{\mu\nu}$ is the metric tensor field on $M$, and $\Phi$ represents the matter fields of the theory. The DPMs of GR are the subset of the KPMs that obey the Einstein equation, which is given by
\begin{equation}
G_{\mu\nu} = 8\pi T_{\mu\nu},
\end{equation}
where
\begin{equation}
G_{\mu\nu} := R_{\mu\nu} - \frac{1}{2}Rg_{\mu\nu}
\end{equation}
is the familiar Einstein tensor and $T_{\mu\nu}$ is the stress-energy tensor. For Knox, the Einstein-Hilbert action contains all of these objects in which we are interested and which comprise the kinematic possibilities of GR; varying this action isolates the dynamical possibilities. We could likewise identify TPG with the KPMs $\left<M, e^a_\mu, \Phi\right>$, whose DPMs are the subset of KPMs that also obey the Einstein field equations (written in terms of the primitive objects of TPG, i.e.~the objects specified in the KPMs of that theory).

Read through this lens, both Weatherall and Knox are operating within a framework whereby they are identifying the relevant empirical content of the theories they are interested in with the dynamics obeyed by the models that comprise these theories. It is a very straightforward argument. There are theories given by models of the form $ \left<M, g_{\mu\nu}, \Phi\right>$ and $\left<M, e^a_\mu, \Phi\right>$, as well as $\left<M, \eta_{\mu\nu}, F_{\mu\nu}, J^\mu \right>$ and $\left<M, \eta_{\mu\nu}, A_{\mu}, J^\mu \right>$. The first pair obeys the dynamics encoded by the Einstein field equations and the second pair obeys the dynamics encoded by the Maxwell equations. Therefore, both pairs are empirically equivalent to each other. The key assumption, of course, is that dynamics is sufficient to fully specify the empirical content of these theories and the models that comprise them. Yet, as we have already seen, there is important empirical content that this characterization leaves out: namely, the empirical content associated with boundary conditions and boundary-related phenomena.

\subsection{Boundary Possible Models}\label{BPM}
In both of the examples considered, boundary conditions play a crucial role in determining the empirical content of models derived from the respective theories, particularly as it relates to describing subsystems. When the empirical information within these models is cast exclusively in terms of dynamics as in the standard semantic view, this additional empirical information is not accounted for in adjudications of theory equivalence, leading to sometimes bizarre and counter-intuitive results when these arguments are taken at face value. The above examples offer helpful illustrations of the importance of boundary conditions in empirical claims, and mesh well with recent philosophical investigations concerning the role of boundary conditions in scientific inquiry. In particular, \textcite{Bursten2021-BURTFO-13} argues that while boundary conditions have traditionally been understood in the philosophy literature as contingent facts akin to initial conditions, they are more properly understood as components of mathematical models. This has to do with, among other things, their role in specifying the scope of mathematical models and generating descriptions of novel phenomena. In a similar spirit, \textcite{Travis} emphasizes that boundary conditions sometimes display behavior and admit of generalizations most often associated with laws.

\begin{figure}
    \centering
    \includegraphics[scale=0.4]{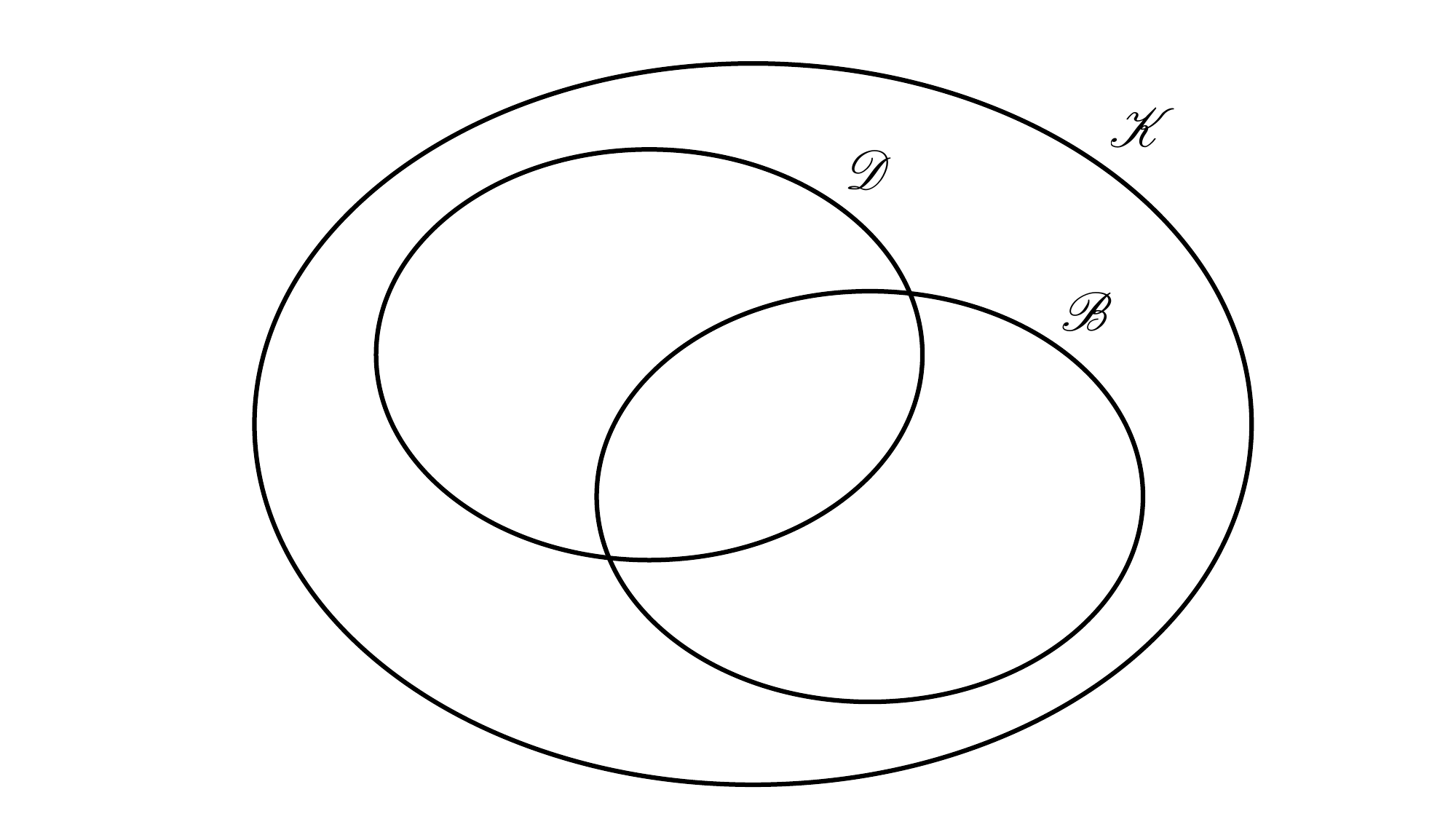}
    %\caption{Caption}
    \caption{The relation between $\mathcal{K}$,$\mathcal{D}$, and $\mathcal{B}$ for a generic $\mathcal{T}$.}
    \label{fig:my_label}
\end{figure}

Considering both the specific examples we have discussed and these more general observations, this suggests a possible modification of the now-standard KPM/DPM version of the semantic approach to account for the role of boundary conditions in the models that are taken to capture the structure of our scientific theories. Here, we introduce a third class of models---proposed by \textcite{ReadBPhil}---known as `boundary possible models' $\mathcal{B}$ (BPMs). Here, $\mathcal{B} \subset \mathcal{K}$, and would denote the subset of KPMs compatible with particular boundary conditions. Then, those $\mathcal{B} \cap \mathcal{D} \subset \mathcal{K}$ would specify those KPMs that are compatible with both particular boundary conditions and particular dynamics. This is depicted in Figure \ref{fig:my_label}.\footnote{In this figure and the surrounding discussion, we are now envisaging including boundary conditions explicitly in the models.}

How could this help us spell out the empirical equivalence of say, EM$_1$ and EM$_2$, in a way that captures this richer view of empirical content? For EM$_1$ and EM$_2$, this is very straightforward. We listed the boundary conditions relevant to describing the boundary surface of a perfect conductor. Let us call them $\mathcal{B}_C$. Both models from EM$_1$ and EM$_2$ have dynamics given by Maxwell's equations. Let us call them $\mathcal{D}_M$. The empirical content of the model from EM$_1$ is then given by the subset of KPMs defined by $\mathcal{B}_{C} \cap \mathcal{D}_{M} $ and the empirical content of the model from EM$_2$ is likewise given by $\mathcal{B}_{C} \cap \mathcal{D}_{M} $. Clearly, these models now possess the same empirical information because we have included the boundary conditions necessary to pick out a unique description of the subsystem within the philosophical criteria that dictate the structural content of the models. Recall that this was previously unavailable in EM$_1$ when its models were only described in terms of dynamics. 

The key is simply to realize that boundary conditions are essential information in any attempt to represent a subsystem-environment decomposition. That is, whether we are using EM$_1$ or EM$_2$, we \textit{must} specify boundary conditions in order to actually build solutions for the mathematical objects of which those descriptions make use (electric and magnetic fields versus gauge fields, respectively). Yet, under the standard semantic way of expressing these theories (in terms of mathematical objects and their dynamics), for a system like the Faraday cage one formulation contains more empirical information than the other precisely because the boundary conditions we used in building the mathematical objects are left out of the formal description of the theory. To be clear, this information is readily available (and often models cannot even be constructed without it) any time one uses standard techniques in electromagnetism. What we are pointing out is that this information has not found its way into the philosophical criteria we use to describe the content of the theory. If boundary conditions are admitted to the formal criteria that defines the structure of a theory, this incongruity dissolves because these boundary conditions contain the information that is needed for the Faraday tensor formulation EM$_1$ to distinguish between different surface charges from within the Faraday cage; something that EM$_2$ more naturally does because information regarding the surface charges finds its way into the gauge potentials. Thus, it then becomes clear that EM$_1$ and EM$_2$ are indeed empirically equivalent once we admit boundary conditions into the semantic criteria. Of course, one could still resist the claim that they are theoretically equivalent due to interpretational issues and implications stemming from the different ontologies postulated by EM$_1$ and EM$_2$ (e.g.\ see  \textcite{Maudlin2018} or \textcite{Teitel2021-TEIWTE}), but at least their empirical equivalence in this context is secure.

%In this simple example, it is easy to see that we can understand both the Faraday tensor and gauge field formulations as empirically equivalent if we think of the boundary conditions listed above as empirical information that is relevant to such a determination. The point of this exercise is to demonstrate that this particular empirical information is not available in the dynamics of the theory. Here, the attempt to demonstrate theoretical equivalence gets tripped up, not in the nuanced technicalities surrounding formal notions of equivalent, but rather in the more mundane issue of empirical equivalence. We can indeed argue that these theories are empirically equivalent and restore our very plausible intuitions that they are equivalent full stop, but in some cases we must add further empirical information that goes beyond dynamics in order to secure this verdict. 

Coming back to the more complicated example of TPG and GR, the action $S_{TPG}$ and its variation $\delta S_{TPG} = 0$ captures the empirical content for models that are compatible with both Dirichlet boundary conditions and the Einstein field equations. That is $S_{TPG}$ gives us the subset of the KPMs that satisfies Dirichlet boundary conditions and the dynamics of the Einstein field equations $\mathcal{B}_{D} \cap \mathcal{D}_{EFE} $. As we saw, while the Einstein-Hilbert action (which we will now switch to specifying as $S_{EH}$) shares the same dynamics $\mathcal{D}_{EFE}$, it is not capable of representing isolated subsystems with the Dirichlet boundary conditions $\mathcal{B}_{D}$. This invites the question: can the models derived from the Einstein-Hilbert action represent any isolated subsystems and can isolated subsystems with Dirichlet boundary conditions be modeled within the framework of GR at all?

The answer to the former question is that there are boundary conditions that make the Einstein-Hilbert action well-defined. Recall that the normal derivatives of the metric did not vanish when examining the boundary terms in \eqref{EHboundary}. Neumann boundary conditions, rather than specifying the values of the metric on the boundary, specify the values of the metric's derivatives on the boundary. It turns out that when one imposes suitable Neumann boundary conditions, both the terms involving tangential and normal derivatives with respect to the metric vanish \parencite{Freidel:2021cjp}. We can then clearly see with this framework that $S_{EH}$ and  $S_{TPG}$ do not share the same empirical content because $ \mathcal{B}_{N} \cap \mathcal{D}_{EFE} \neq \mathcal{B}_{D} \cap \mathcal{D}_{EFE}$ (recall again that $S_{TPG}$ gives us the subset of the KPMs $\mathcal{B}_{D} \cap \mathcal{D}_{EFE}$).
%\todo{This isn't clear, because TPG isn't mentioned anywhere in the condition.}

Finally, how does GR actually model isolated subsystems with Dirichlet boundary conditions and study important concepts found in asymptotic spacetimes? The answer is that we must set aside the Einstein-Hilbert action $S_{EH}$ in favor of what is known as the Gibbons-Hawking-York (GHY) action $S_{GHY}$:
\begin{equation}
    S_{\mathrm{GHY}}=\frac{1}{16 \pi G} \int d^{4} x \sqrt{-g} R + \frac{1}{8 \pi G} \oint_{\partial M} d^{3} \Omega \epsilon \sqrt{h} K,
\end{equation}
where $K=\nabla^\mu n_\mu$ is the trace of the extrinsic curvature, $h$ is the induced metric on the boundary, and $\epsilon$ is $+1$ when the boundary hypersurface is spacelike and $-1$ when the boundary hypersurface is timelike \parencite{York:1972sj, Gibbons:1976ue}. We see here that this action is equal to the Einstein-Hilbert action plus a boundary term. When varying this action, we find the bulk term that contains the dynamical Einstein field equations $G_{\mu\nu}$, the boundary term from before, and a further boundary term originating from the GHY term. Upon imposing Dirichlet boundary conditions $\left.\delta g_{\mu \nu}\right|_{\partial M}=0$, we find that the variation of the GHY boundary term exactly cancels out the previously non-vanishing terms. Thus, in the presence of manifolds with boundaries with Dirichlet boundary conditions, we have
\begin{equation}
\delta S_{\mathrm{GHY}}=\frac{1}{16 \pi G} \int_{M} d^{4} x \sqrt{-g} G_{\mu \nu} \delta g^{\mu \nu}.
\end{equation}
This follows the exact same pattern as the variation of the TPG action. The additional boundary term plays a similar role and cancels out previously problematic terms, yielding a well-defined variation. 

We see that $S_{GHY}$ gives us the subset of KPMs $\mathcal{B}_{D} \cap \mathcal{D}_{EFE}$. This matches up with the subset of KPMs given to us by $S_{TPG}$, which as we have seen is also $\mathcal{B}_{D} \cap \mathcal{D}_{EFE}$. Indeed, $\delta S_{TPG} = \delta S_{GHY}$ when $\mathcal{B}_{D}$ is imposed, so we know that both actions share the same dynamical content and the same representational capacity when it comes to isolated subsystems. Important quantities that depend on these boundary terms and conditions such as the ADM mass $M_{ADM}$ and black hole entropy $S_{BH}$ are found to be in agreement. For example, $M_{ADM}$ is one of the quantities to which Penrose referred and represents the mass-energy content of a spacetime. Using $S_{GHY}$ and $S_{TPG}$ to determine this quantity gives the same results, which are crucially dependent on the role and behavior of the boundary terms and conditions that we have discussed \parencite{Dyer:2008hb, Wald:1993nt, Iyer:1994ys, Hammad:2019oyb}. As \textcite{Freidel:2021bmc} have noted, these boundary terms can also effectively bring the Noether charges of a theory into alignment with the corresponding Hamiltonian charges (i.e.,~the ADM mass), which connects such quantities to Hamiltonian observables. Coming to black hole entropy $S_{BH}$, one can use the Euclidean semi-classical path integral approach and find that one obtains identical results for this quantity, with the boundary terms present in both $S_{TPG}$ and $S_{GHY}$ contributing the entire entropy in the calculation \parencite{Gibbons:1976ue, Gibbons:1978ac, Oshita:2017nhn}.

How could one go about arguing for theoretical equivalence of GR and TPG given our characterization of the semantic view that includes KPMs, DPMs, and BPMs? One way would involve taking inspiration from the characterization of equivalence found in \textcite{Nguyen2017-NGUSRA}, which he has dubbed `representational equivalence'. This would mean showing that models from both GR and TPG can represent the same target systems, and that they make the same empirical claims about these target systems. We have already partially done that by showing $S_{GHY}$ and $S_{TPG}$ coincide in the target subsystems they can represent and discussing how they align in the empirical claims they make about boundary dependent phenomena that goes beyond the shared dynamics of all these models. One could similarly investigate other actions, models, and isolated subsystems in both GR and TPG and ensure that they align in both representational capacity and empirical claims. This still leaves open the admittedly more difficult interpretative questions regarding whether GR and TPG license all of the same interpretive claims about the world and their target systems, but it at least provides a straightforward path to demonstrating perspicuously their empirical equivalence. 

Our conception of a theory should specify the empirical content of the theory. KPMs define the objects of interest to us within a particular theory, but we would not say that defining a theory exclusively in terms of KPMs is satisfying because it plainly fails to specify empirical content. We also want to specify how these objects interact with each other and behave empirically. DPMs specify their dynamics. However, as the above examples demonstrate, dynamics does not constitute the full extent of the empirical content of these models. We also want to specify the subsystem-environment decompositions that these models can represent, as well as any boundary related empirical content that goes beyond the dynamics of these objects. Just as KPMs are insufficient to fully specify a theory's empirical content, so too are DPMs alone: the latter should be supplemented with BPMs to more fully specify to empirical content of a theory. 

Coming back to the issues of empirical and theoretical equivalence, it is clear that one's conception of theory content and structure will have a non-trivial impact on any subsequent adjudication of theoretical equivalence. The identification of GR's content with the dynamics resulting from the Einstein-Hilbert action and of EM$_1$'s content with a Faraday tensor obeying Maxwell's equations does not fully specify the empirical content of those theories, and thus is responsible for incorrect adjudications of empirical equivalence when compared with their allegedly equivalent counterparts. Both Knox and Weatherall do make some qualifying statements. \textcite[p.~272]{Knox2011-KNONTA} notes that the local equivalence of the TPG and EH actions up to a divergence may lead to some global worries, while \textcite[p.~1078]{Weatherall2016-WEAANG} notes that he stipulates that the empirical content of EM is exhausted by Faraday tensors compatible with Maxwell's equations. Yet, it is clear that in both cases, there are indeed global worries that render their adjudications problematic and that these qualifying statements do not do justice to the empirical content that is lost when one looks exclusively at local dynamics. Overall, as we have argued above, their analyses and conclusions can still obtain provided that these further considerations are accounted for. However, it is important to acknowledge both that these models require additional specifications beyond the equations of motion in order to generate the totality of their empirical content and that this is a relevant consideration in adjudicating equivalence.

\subsection{The Pragmatic View} \label{Pragmatic}

We can also draw from the pragmatic view of theories to illuminate these adjudications of theoretical equivalence as well as the importance of considering carefully one's view of theory structure. Rather than totally repudiating the syntactic and semantic views, the pragmatic view acknowledges the utility of many of the formal components of these other perspectives, while also emphasizing non-formal considerations. While there is significant variety amongst proponents of this view \parencite{Cartwright1983-CARHTL, Hacking1983-HACRAI-7, Kitcher1993-KITTAO-2, sep-structure-scientific-theories}, two strands of thought stand out as particularly relevant to the present discussion: (i) model pluralism and (ii) focus on scientific practice.

On (i): Cartwright claims that models are the appropriate level of scientific investigations (as opposed to theories) and argues that there are many different but legitimate reasons to utilize different models. ``Models serve a variety of purposes, and individual models are to be judged according to how well they serve the purpose at hand" \parencite[p.~152]{Cartwright1983-CARHTL}. One model might be focused on accuracy for a particular quantity, while another might be trying to incorporate additional phenomena into the description and consequently, might be less focused on maximizing accuracy of any one particular quantity. 

This point is made quite generally, but we can see something similar going on in GR. We have already encountered two actions used in GR, $S_{EH}$ and $S_{GHY}$, but there are others, including but not limited to the $\Gamma$-$\Gamma$ action
\begin{equation}
S_{\Gamma\Gamma} = \frac{1}{16 \pi G} \int d^4 x\sqrt{-g} g^{\mu \nu}\left(\Gamma_{\mu \beta}^{\alpha} \Gamma_{\alpha \nu}^{\beta}-\Gamma_{\mu \nu}^{\alpha} \Gamma_{\alpha \beta}^{\beta}\right)
\end{equation}
and the ADM action
\begin{equation}
S_{ADM} = \frac{1}{16 \pi G} \int d^4 x\sqrt{-g} (\Tilde{R} + K^{\mu\nu}K_{\mu\nu} - K^2),
\end{equation}
where $\Tilde{R}$ is the three-dimensional Ricci scalar of the spatial slice in the $3+1$ decomposition in the $ADM$ formulation and $K$ is the extrinsic curvature. $S_{\Gamma\Gamma}$ turns out to be incredibly convenient for demonstrating that GR corresponds to the self-coupling of a massless spin-2 particle, due to the cubic nature of the form of the Lagrangian, which is in analogy with both Yang-Mills fields and spin-1 particles and chiral fields and spin-zero particles \parencite{Deser:1969wk, Deser_1987}. Additionally, the $3+1$ decomposition like that used in $S_{ADM}$ is particularly important because, among many other benefits, it is  useful for solving initial value problems as it allows us to mathematically formulate the Einstein equations as ``a Cauchy problem with constraints" \parencite[pp.~11--12]{Gourgoulhon:2007ue}. Consequently, it has become the foundation for most approaches in numerical relativity. Whether we choose an action based on convenience, clarity, or necessity, there are a lot of options at our disposal for modeling phenomena in GR. 
%If we view theories in purely formal terms based solely on the dynamical content of their models, there is nothing problematic with identifying GR with the dynamical content of the Einstein-Hilbert action. But 
Under this pragmatic approach of embracing model pluralism, it is clear that GR is much broader than the dynamical content of one of these actions alone and that any adjudication of theoretical equivalence would need to address this broader scope.

Another theme that the pragmatic view emphasizes---point (ii) above---is that our view of theories should be commensurate with scientific practice. While acknowledging the utility of formal criteria, \textcite[p.~7]{NicTeh2022} has argued that a theory should be more properly viewed as a collection of physical representations, ``accompanied by a keen `know how’ about what we can do with such representations and how they are related to each other." This emphasis on `know how' implores us to consider scientific practice in specifying the structure of theories and has indeed been a major focus of advocates for the pragmatic view \parencite{Hacking1983-HACRAI-7, Kitcher1993-KITTAO-2}. Clearly, practitioners of GR use many different dynamically equivalent actions depending on the problem at hand, but this discussion also highlights how boundary phenomena have become more relevant in both physics and philosophy communities in recent years. As we have already noted, physicists have been exploring boundary phenomena and isolated subsystems, with examples including edge modes in the quantum Hall effect, black hole entropy, and slip/no-slip boundary conditions in fluid flow, while philosophers have been interested in them as a way to cash out the direct empirical significance of symmetries and the explanatory capabilities of models. Furthermore, it is worth emphasizing that these examples in physics feature novel phenomena that are apparent only when we consider boundary content as descriptions of these phenomena are not available from bulk dynamics alone. Consequently, our views on theory structure should be updated to accommodate these kinds of empirical phenomena.

Here, we see a potential connection between the semantic and pragmatic approaches. While the pragmatic view does emphasize non-formal elements of modeling and theory structure, its embrace of pluralism also allows it to accommodate a variety of strategies in describing theory structure, including the use of more formal notions. Indeed, some philosophers have even argued that ``the semantic conception in its bare minimal expression" is very compatible with ``pragmatic elements and themes" \parencite[p.~348]{Pero}. We can thus rely on pragmatic considerations such as scientific practice to inform us of what structures should find their way into a formal representations of the models in our theories. Before the theoretical and empirical importance of boundaries was truly appreciated, it might have made more sense to view a theory exclusively in terms of its dynamics and mathematical objects. However, as scientific practice (and philosophical interest) has changed and brought this boundary phenomena more into focus, it now makes sense to adjust our views on the structure of theories to be commensurate with scientific practice. As we saw in the previous section, one can easily accommodate boundary conditions within a tradition semantic analysis of a theory.

\section{Consequences and Conclusions}

\noindent Discussions concerning both the equivalence and structure of physical theories have been and will continue to be important themes in the philosophy of science. As we have seen (following \textcite{Barrett2019-BAREAI-10}), each of these questions bears upon the other because adopting a particular standard of equivalence will necessarily specify a view of what the contentful features of a theory actually are; and similarly, adopting a particular view of theory content or structure will necessarily set the standard by which equivalence is to be judged. 

The aforementioned examples in the literature regarding the supposed theoretical equivalence between EM$_1$ and EM$_2$ and between TPG and GR illustrate both that these questions do indeed interact with each other and suggest that these questions need to be tackled in parallel. In navigating these issues surrounding theory equivalence and structure, we take one moral from this discussion to be that adopting a pragmatic attitude towards theory structure can be very fruitful. Indeed, we saw that in both examples considered, the source of the failure of empirical equivalence came about from the authors adopting views of theory structure that, while useful and consistent with a fairly standard view the philosophy literature, used formal criteria that were overly restrictive regarding the empirical substructures that one could attribute to the theories. Thus, additional empirical content related to boundary phenomena and isolated subsystems did not make its way into the analysis.

However, the pragmatic view can help bring these discussions of equivalence and structure into alignment. As we have seen in these examples, the pragmatic view indicates that we should be pluralistic regarding our representations of models and theories, as well as update the components we consider when utilizing formal descriptions of theory structure by supplementing the standard semantic representation with boundary conditions. In so doing, we can construct an argument for the equivalence of TPG and GR that also reflects the full richness of the empirical content that these theories are currently understood to possess. While the example of EM$_1$ and EM$_2$ is not quite as dramatic given that the interpretational issues are not generally considered to be quite as difficult, there is something similar going on. When boundary conditions are included in the formal criteria that describe theory structure, it is clear that EM$_1$ and EM$_2$ are equivalent and that the issue merely stemmed from adopting an overly restrictive view of the empirical content contained within the formal descriptions. Furthermore, this pragmatic attitude provides flexibility in that it allows us to continuously update our understanding of theory structure as previous empirical substructures become better understood and novel empirical substructures come into view. In the context of these many empirical realizations surrounding boundary phenomena and their increased importance to both physicists and philosophers, it is clear that such an update is needed and that boundary conditions and phenomena must be considered in discussions of theory structure and equivalence.

\section*{Acknowledgements}

\noindent We are grateful to Caspar Jacobs, Travis McKenna, Ruward Mulder, Mauricio Suárez, Nic Teh, and Jim Weatherall for helpful discussions. This work was supported by grant number 61521 from the John Templeton Foundation.

\begin{comment}
------

On the intersection of the boundary stuff and the pragmatics stuff: I take the essential point of the former to be that whether theories come out as equivalent depends upon what you want to preserve; if you include boundary objects, theories such as GR and TPG don't come out as equivalent. I take the essential point of the pragmatics section to be that it might be better to be pluralist about the kinds of actions which can constitute a theory anyway.

How do these interact? Well, if one embraces the first point but not the second (the second being pluralism), then TPG and GR aren't equivalent. If one embraces neither, then TPG and GR are equivalent (a la Knox). If one embraces both, then TPG and GR in a sense become equivalent again, since the TPG action is equivalent to the ADM action, and the latter falls within the scope of GR given pluralism.

Do you agree? And if so how would it go in the case of EM? One main different is that, unlike the GR case (with the ADM action etc.), there's no alternative F-field action. But: two theories are equivalent when one puts in a bounary condition for the F-field formulation by hand.
%So even a pluralist attitude doesn't warrant the verdict that these two approaches to

\end{comment}
    
\printbibliography

@article{Wolf:2023rad,
    author = "Wolf, William J. and Read, James",
    title = "{The Non-Relativistic Geometric Trinity of Gravity}",
    eprint = "2308.07100",
    archivePrefix = "arXiv",
    primaryClass = "gr-qc",
    year = "2023"
}

@article{Glymour1970-GLYTRA,
	title = {Theoretical Realism and Theoretical Equivalence},
	year = {1970},
	author = {Glymour, Clark},
	journal = {PSA: Proceedings of the Biennial Meeting of the Philosophy of Science Association},
	pages = {275--288},
	volume = {1970},
	publisher = {University of Chicago Press}
}

@article{Quine1975-QUIOEE,
	publisher = {Springer},
	doi = {10.1007/bf00178004},
	title = {On Empirically Equivalent Systems of the World},
	pages = {313--28},
	number = {3},
	author = {Willard Quine},
	volume = {9},
	year = {1975},
	journal = {Erkenntnis}
}

@unpublished{Weatherall2018-WEATEI,
	author = {Weatherall, James},
	year = {2018},
	title = {Theoretical Equivalence in Physics},
    eprint={1810.08192},
    archivePrefix={arXiv}
}

@book{vonNeumann+2018,
author = {{von Neumann}, John},
editor = {Nicholas A. Wheeler},
doi = {doi:10.1515/9781400889921},
title = {Mathematical Foundations of Quantum Mechanics: New Edition},
year = {2018},
publisher = {Princeton University Press},
ISBN = {9781400889921},
lastchecked = {2022-07-01}
}

@article{Muller1997-MULTEM,
	doi = {10.1016/s1355-2198(96)00022-6},
	number = {1},
	author = {F. A. Muller},
	journal = {Studies in History and Philosophy of Science Part B: Studies in History and Philosophy of Modern Physics},
	publisher = {Elsevier},
	year = {1997},
	title = {The Equivalence Myth of Quantum Mechanics --Part I},
	pages = {35--61},
	volume = {28}
}

@article{Muller1997-MULTEM-2,
	pages = {219--247},
	number = {2},
	author = {F. A. Muller},
	title = {The Equivalence Myth of Quantum Mechanics--Part II},
	journal = {Studies in History and Philosophy of Science Part B: Studies in History and Philosophy of Modern Physics},
	year = {1997},
	doi = {10.1016/s1355-2198(97)00001-4},
	volume = {28}
}

@article{Dyson,
  title = {The Radiation Theories of Tomonaga, Schwinger, and Feynman},
  author = {Dyson, F. J.},
  journal = {Phys. Rev.},
  volume = {75},
  issue = {3},
  pages = {486--502},
  numpages = {0},
  year = {1949},
  publisher = {American Physical Society},
}

@article{Barrett2019-BAREAI-10,
	number = {4},
	pages = {1167--1199},
	volume = {70},
	journal = {British Journal for the Philosophy of Science},
	year = {2019},
	author = {Thomas William Barrett},
	title = {Equivalent and Inequivalent Formulations of Classical Mechanics},
	doi = {10.1093/bjps/axy017}
}

@article{Curiel,
author = {Curiel, Erik},
title = {Classical Mechanics Is Lagrangian; It Is Not Hamiltonian},
journal = {The British Journal for the Philosophy of Science},
volume = {65},
number = {2},
pages = {269-321},
year = {2014},
}

@article{North2009-NORTSO-14,
	author = {Jill North},
	title = {The Structure of Physics: A Case Study},
	volume = {106},
	journal = {Journal of Philosophy},
	doi = {jphil2009106213},
	publisher = {Journal of Philosophy},
	pages = {57--88},
	number = {2},
	year = {2009},
}

@article{Maldacena:1997re,
    author = "Maldacena, Juan Martin",
    title = "{The Large N limit of superconformal field theories and supergravity}",
    eprint = "hep-th/9711200",
    archivePrefix = "arXiv",
    reportNumber = "HUTP-97-A097, HUTP-98-A097",
    doi = "10.1023/A:1026654312961",
    journal = "Adv. Theor. Math. Phys.",
    volume = "2",
    pages = "231--252",
    year = "1998"
}

@article{deHaro2016-DEHCAO-2,
	pages = {1381--1425},
	year = {2016},
	publisher = {Springer},
	doi = {10.1007/s10701-016-0037-4},
	number = {11},
	title = {Conceptual Aspects of Gauge/Gravity Duality},
	volume = {46},
	journal = {Foundations of Physics},
	author = {De Haro, Sebastian and Daniel Mayerson and Jeremy Butterfield}
}

@incollection{VANFRAASSEN1986307,
title = {Aim and Structure of Scientific Theories},
editor = {Ruth {Barcan Marcus} and Georg J.W. Dorn and Paul Weingartner},
series = {Studies in Logic and the Foundations of Mathematics},
publisher = {Elsevier},
volume = {114},
pages = {307-318},
year = {1986},
booktitle = {Logic, Methodology and Philosophy of Science VII},
author = {Bas C. {van Fraassen}},
}

@article{Knox2011-KNONTA,
	author = {Eleanor Knox},
	number = {4},
	title = {Newton-Cartan Theory and Teleparallel Gravity: The Force of a Formulation},
	doi = {10.1016/j.shpsb.2011.09.003},
	volume = {42},
	pages = {264--275},
	publisher = {Elsevier},
	journal = {Studies in History and Philosophy of Science Part B: Studies in History and Philosophy of Modern Physics},
	year = {2011}
}

@article{Weatherall2016-WEAANG,
	title = {Are Newtonian Gravitation and Geometrized Newtonian Gravitation Theoretically Equivalent?},
	author = {Weatherall, James},
	publisher = {Springer Verlag},
	pages = {1073--1091},
	journal = {Erkenntnis},
	volume = {81},
	doi = {10.1007/s10670-015-9783-5},
	year = {2016},
	number = {5}
}

@article{WallaceFields,
    author = "Wallace, David",
    title = "{Fields as Bodies: a unified presentation of spacetime and internal gauge symmetry}",
    eprint = "1502.06539",
    archivePrefix = "arXiv",
    primaryClass = "gr-qc",
    year = "2015"
}

@book{ReadBPhil,
author = {Read, James},
year = {2016},
title = {Background independence in classical and quantum gravity},
note = {B.Phil. Thesis},
publisher = {University of Oxford}}

@article{MR,
	number = {2},
	pages = {229--251},
	year = {2021},
	author = {James Read and Tushar Menon},
	publisher = {Springer Verlag},
	title = {The Limitations of Inertial Frame Spacetime Functionalism},
	volume = {199},
	doi = {10.1007/s11229-019-02299-2},
	journal = {Synthese}
}

@article{Coffey,
author = {Coffey, Kevin},
title = {Theoretical Equivalence as Interpretative Equivalence},
journal = {The British Journal for the Philosophy of Science},
volume = {65},
number = {4},
pages = {821-844},
year = {2014},
doi = {10.1093/bjps/axt034},
eprint = { 
        https://doi.org/10.1093/bjps/axt034
    
}
}

@article{Wallace2014-GREECO,
	number = {1},
	journal = {British Journal for the Philosophy of Science},
	volume = {65},
	year = {2014},
	author = {D. Wallace and Hilary Greaves},
	pages = {59--89},
	publisher = {Oxford University Press},
	title = {Empirical Consequences of Symmetries},
	doi = {10.1093/bjps/axt005}
}

@article{Teh2016-TEHGGU,
	title = {Galileo's Gauge: Understanding the Empirical Significance of Gauge Symmetry},
	author = {Nicholas J. Teh},
	number = {1},
	pages = {93--118},
	publisher = {University of Chicago},
	volume = {83},
	doi = {10.1086/684196},
	year = {2016},
	journal = {Philosophy of Science}
}

@article{MurgueitioRamirezForthcoming-MURAGS-3,
	author = {Murgueitio Ram\'{i}rez, Sebasti\'{a}n and Teh, Nicholas},
	year = {forthcoming},
	journal = {British Journal for the Philosophy of Science},
	title = {Abandoning Galileo's Ship: The Quest for Non-Relational Empirical Significance}
}

@article{Gomes2021-GOMHAT,
	title = {Holism as the Empirical Significance of Symmetries},
	author = {Henrique Gomes},
	number = {3},
	year = {2021},
	volume = {11},
	doi = {10.1007/s13194-021-00397-y},
	pages = {1--41},
	journal = {European Journal for Philosophy of Science},
	publisher = {Springer Verlag}
}

@article{Wolf:2021ydy,
    author = "Wolf, William J. and Read, James and Teh, Nicholas J.",
    title = "{Edge Modes and Dressing Fields for the Newton\textendash{}Cartan Quantum Hall Effect}",
    eprint = "2111.08052",
    archivePrefix = "arXiv",
    primaryClass = "cond-mat.mes-hall",
    doi = "10.1007/s10701-022-00615-4",
    journal = "Found. Phys.",
    volume = "53",
    number = "1",
    pages = "3",
    year = "2023"
}

@article{Bursten2021-BURTFO-13,
	volume = {88},
	journal = {Philosophy of Science},
	doi = {10.1086/711502},
	number = {2},
	year = {2021},
	author = {Julia R. S. Bursten},
	title = {The Function of Boundary Conditions in the Physical Sciences},
	pages = {234--257}
}

@article{Wen:1995qn,
    author = "Wen, Xiao-Gang",
    title = "{Topological orders and edge excitations in FQH states}",
    eprint = "cond-mat/9506066",
    archivePrefix = "arXiv",
    reportNumber = "PRINT-95-148 (MIT)",
    doi = "10.1080/00018739500101566",
    journal = "Adv. Phys.",
    volume = "44",
    number = "5",
    pages = "405--473",
    year = "1995"
}

@article{Gibbons:1976ue,
    author = "Gibbons, G. W. and Hawking, S. W.",
    title = "{Action Integrals and Partition Functions in Quantum Gravity}",
    reportNumber = "PRINT-76-0995 (CAMBRIDGE)",
    doi = "10.1103/PhysRevD.15.2752",
    journal = "Phys. Rev. D",
    volume = "15",
    pages = "2752--2756",
    year = "1977"
}

@book{VanFraassenBas1980-VANTSI,
	publisher = {Oxford, England: Oxford University Press},
	title = {The Scientific Image},
	year = {1980},
	author = {{van Fraassen}, Bas C.}
}

@article{Barrett2016-BARME-5,
	journal = {Review of Symbolic Logic},
	publisher = {Cambridge University Press (Cup)},
	year = {2016},
	author = {Thomas William Barrett and Hans Halvorson},
	doi = {10.1017/s1755020316000186},
	title = {Morita Equivalence},
	number = {3},
	volume = {9},
	pages = {556--582}
}

@unpublished{CurielKinematics,
	year = {2016},
	title = {Kinematics, Dynamics, and the Structure of Physical Theory},
	author = {Erik Curiel}
}

@misc{curiel2020schematizing,
      title={Schematizing the Observer and the Epistemic Content of Theories}, 
      author={Erik Curiel},
      year={2020},
      eprint={1903.02182},
      archivePrefix={arXiv},
      primaryClass={physics.hist-ph}
}

@article{BradleyWeatherall,
	author = {Clara Bradley and James Owen Weatherall},
	year = {2020},
	volume = {50},
	number = {4},
	doi = {10.1007/s10701-020-00330-y},
	pages = {270--293},
	publisher = {Springer Us},
	title = {On Representational Redundancy, Surplus Structure, and the Hole Argument},
	journal = {Foundations of Physics}
}

@article{LinnemannReadKin,
	author = {Linnemann, Niels and Read, James},
	doi = {10.1007/s10701-021-00453-w},
	journal = {Foundations of Physics},
	number = {2},
	pages = {53},
	title = {On the Status of Newtonian Gravitational Radiation},
	url = {https://doi.org/10.1007/s10701-021-00453-w},
	volume = {51},
	year = {2021}
}

@article{Travis,
	journal = {British Journal for the Philosophy of Science},
	publisher = {},
	year = {forthcoming},
	author = {McKenna, Travis},
	doi = {10.1086/725906},
	title = {Laws of Nature and Their Supporting Casts},
	number = {},
	volume = {},
	pages = {}
}

@book{Aldrovandi:2013wha,
    author = "Aldrovandi, Ruben and Pereira, Jos\'e Geraldo",
    title = "{Teleparallel Gravity}: {An Introduction}",
    doi = "10.1007/978-94-007-5143-9",
    %isbn = "978-94-007-5142-2, 978-94-007-5143-9",
    publisher = "Springer",
    year = "2013"
}

@article{Gourgoulhon:2007ue,
    author = "Gourgoulhon, Eric",
    title = "{3+1 formalism and bases of numerical relativity}",
    eprint = "gr-qc/0703035",
    archivePrefix = "arXiv",
    year = "2007"
}

@ARTICLE{2005cond.mat..1557L,
       author = {{Lauga}, Eric and {Brenner}, Michael P. and {Stone}, Howard A.},
        title = "{Microfluidics: The no-slip boundary condition}",
      journal = {arXiv e-prints},
         year = 2005,
          eid = {cond-mat/0501557},
        pages = {cond-mat/0501557},
          doi = {10.48550/arXiv.cond-mat/0501557},
archivePrefix = {arXiv},
       eprint = {cond-mat/0501557},
 primaryClass = {cond-mat.soft},
       adsurl = {https://ui.adsabs.harvard.edu/abs/2005cond.mat..1557L}
}

@book{griffiths_2017, place={Cambridge}, edition={4}, title={Introduction to Electrodynamics}, DOI={10.1017/9781108333511}, publisher={Cambridge University Press}, author={Griffiths, David J.}, year={2017}}

@book{zangwill_2012, place={Cambridge}, title={Modern Electrodynamics}, DOI={10.1017/CBO9781139034777}, publisher={Cambridge University Press}, author={Zangwill, Andrew}, year={2012}}

@inbook{Penrose1982SOMEUP,
url = {https://doi.org/10.1515/9781400881918-034},
title = {Some Unsolved Problems in Classical General Relativity},
booktitle = {Seminar on Differential Geometry. (AM-102), Volume 102},
author = {R. Penrose},
editor = {Shing-tung Yau},
publisher = {Princeton University Press},
address = {Princeton},
pages = {631--668},
doi = {doi:10.1515/9781400881918-034},
isbn = {9781400881918},
year = {1982},
lastchecked = {2023-02-15}
}

@article{Belot2018-GORFME,
	publisher = {Wiley-Blackwell},
	year = {2018},
	pages = {946--981},
	author = {Gordon Belot},
	title = {Fifty Million Elvis Fans Can't Be Wrong},
	journal = {No\^{u}s},
	doi = {10.1111/nous.12200}
}

@article{BeltranJimenez:2019esp,
    author = "Jim\'enez, Jose Beltr\'an and Heisenberg, Lavinia and Koivisto, Tomi S.",
    title = "{The Geometrical Trinity of Gravity}",
    eprint = "1903.06830",
    archivePrefix = "arXiv",
    primaryClass = "hep-th",
    doi = "10.3390/universe5070173",
    journal = "Universe",
    volume = "5",
    number = "7",
    pages = "173",
    year = "2019"
}

@article{Wolf:2023xrv,
    author = "Wolf, William J. and Sanchioni, Marco and Read, James",
    title = "{Underdetermination in Classic and Modern Tests of General Relativity}",
    eprint = "2307.10074",
    archivePrefix = "arXiv",
    primaryClass = "physics.hist-ph",
    year = "2023"
}

@article{Blau,
  title={Lecture Notes on General Relativity},
  author={Matthias Blau},
  URL={http://www.blau.itp.unibe.ch/newlecturesGR.pdf}
}

@article{Oshita:2017nhn,
    author = "Oshita, Naritaka and Wu, Yi-Peng",
    title = "{Role of spacetime boundaries in Einstein's other gravity}",
    eprint = "1705.10436",
    archivePrefix = "arXiv",
    primaryClass = "gr-qc",
    doi = "10.1103/PhysRevD.96.044042",
    journal = "Phys. Rev. D",
    volume = "96",
    number = "4",
    pages = "044042",
    year = "2017"
}

@article{Lutz2017-LUTWWT,
	number = {2},
	author = {Sebastian Lutz},
	doi = {10.1111/phpr.12221},
	title = {What Was the Syntax-Semantics Debate in the Philosophy of Science About?},
	journal = {Philosophy and Phenomenological Research},
	pages = {319--352},
	volume = {95},
	year = {2017}
}

@book{Cartwright1983-CARHTL,
	publisher = {Oxford, England: Oxford University Press},
	title = {How the Laws of Physics Lie},
	author = {Nancy Cartwright},
	year = {1983}
}

@book{Kitcher1993-KITTAO-2,
	year = {1993},
	publisher = {Oxford, England: Oxford University Press},
	author = {Philip Kitcher},
	title = {The Advancement of Science: Science Without Legend, Objectivity Without Illusions}
}

@book{Hacking1983-HACRAI-7,
	author = {Ian Hacking},
	year = {1983},
	publisher = {Cambridge University Press},
	title = {Representing and Intervening: Introductory Topics in the Philosophy of Natural Science}
}

@InCollection{sep-structure-scientific-theories,
	author       =	{Winther, Rasmus Grønfeldt},
	title        =	{{The Structure of Scientific Theories}},
	booktitle    =	{The {Stanford} Encyclopedia of Philosophy},
	editor       =	{Edward N. Zalta},
	howpublished =	{\url{https://plato.stanford.edu/archives/spr2021/entries/structure-scientific-theories/}},
	year         =	{2021},
	edition      =	{{S}pring 2021},
	publisher    =	{Metaphysics Research Lab, Stanford University}
}

@incollection{Pooley2013-POOSAR,
	booktitle = {The Oxford Handbook of Philosophy of Physics},
	publisher = {Oxford University Press},
	author = {Oliver Pooley},
	title = {Substantivalist and Relationalist Approaches to Spacetime},
	year = {2013},
	editor = {Robert Batterman}
}

@incollection{Pooley2015-POOBID,
	booktitle = {Towards a Theory of Spacetime Theories},
	year = {2015},
	editor = {Dennis Lehmkuhl},
	author = {Oliver Pooley},
	publisher = {Birkh\"{a}user},
	title = {Background Independence, Diffeomorphism Invariance, and the Meaning of Coordinates}
}

@article{Freidel:2021cjp,
	doi = {10.1007/jhep09(2021)083},
  
	url = {https://doi.org/10.1007%2Fjhep09%282021%29083},
  
	year = 2021,
  
	volume = {2021},
  
	number = {9},
  
	author = {Laurent Freidel and Roberto Oliveri and Daniele Pranzetti and Simone Speziale},
  
	title = {Extended corner symmetry, charge bracket and Einstein's equations},
  
	journal = {Journal of High Energy Physics}
}

@article{York:1972sj,
    author = "York, Jr., James W.",
    title = "{Role of conformal three geometry in the dynamics of gravitation}",
    doi = "10.1103/PhysRevLett.28.1082",
    journal = "Phys. Rev. Lett.",
    volume = "28",
    pages = "1082--1085",
    year = "1972"
}

@article{Dyer:2008hb,
    author = "Dyer, Ethan and Hinterbichler, Kurt",
    title = "{Boundary Terms, Variational Principles and Higher Derivative Modified Gravity}",
    eprint = "0809.4033",
    archivePrefix = "arXiv",
    primaryClass = "gr-qc",
    doi = "10.1103/PhysRevD.79.024028",
    journal = "Phys. Rev. D",
    volume = "79",
    pages = "024028",
    year = "2009"
}

@article{Hammad:2019oyb,
    author = "Hammad, F. and Dijamco, D. and Torres-Rivas, A. and B\'erub\'e, D.",
    title = "{Noether charge and black hole entropy in teleparallel gravity}",
    eprint = "1912.08811",
    archivePrefix = "arXiv",
    primaryClass = "gr-qc",
    doi = "10.1103/PhysRevD.100.124040",
    journal = "Phys. Rev. D",
    volume = "100",
    number = "12",
    pages = "124040",
    year = "2019"
}

@article{Wald:1993nt,
    author = "Wald, Robert M.",
    title = "{Black hole entropy is the Noether charge}",
    eprint = "gr-qc/9307038",
    archivePrefix = "arXiv",
    reportNumber = "EFI-93-42",
    doi = "10.1103/PhysRevD.48.R3427",
    journal = "Phys. Rev. D",
    volume = "48",
    number = "8",
    pages = "R3427--R3431",
    year = "1993"
}

@article{Iyer:1994ys,
    author = "Iyer, Vivek and Wald, Robert M.",
    title = "{Some properties of Noether charge and a proposal for dynamical black hole entropy}",
    eprint = "gr-qc/9403028",
    archivePrefix = "arXiv",
    doi = "10.1103/PhysRevD.50.846",
    journal = "Phys. Rev. D",
    volume = "50",
    pages = "846--864",
    year = "1994"
}

@article{Freidel:2021bmc,
    author = "Freidel, Laurent and Teh, Nicholas",
    title = "{Substantive general covariance and the Einstein-Klein dispute: A Noetherian approach}",
    eprint = "2109.08516",
    archivePrefix = "arXiv",
    primaryClass = "physics.hist-ph",
    year = "2021"
}

@article{Gibbons:1978ac,
    author = "Gibbons, G. W. and Hawking, S. W. and Perry, M. J.",
    title = "{Path Integrals and the Indefiniteness of the Gravitational Action}",
    reportNumber = "PRINT-78-0375 (CAMBRIDGE)",
    doi = "10.1016/0550-3213(78)90161-X",
    journal = "Nucl. Phys. B",
    volume = "138",
    pages = "141--150",
    year = "1978"
}

@article{Deser:1969wk,
    author = "Deser, Stanley",
    title = "{Selfinteraction and gauge invariance}",
    eprint = "gr-qc/0411023",
    archivePrefix = "arXiv",
    doi = "10.1007/BF00759198",
    journal = "Gen. Rel. Grav.",
    volume = "1",
    pages = "9--18",
    year = "1970"
}

@article{Deser_1987,
	doi = {10.1088/0264-9381/4/4/006},
	year = 1987,
	publisher = {{IOP} Publishing},
	volume = {4},
	number = {4},
	pages = {L99--L105},
	author = {S Deser},
	title = {Gravity from self-interaction in a curved background},
	journal = {Classical and Quantum Gravity},
}

@article{Pero,
author = {Su\'{a}rez, Mauricio and Pero, Francesca},
title = {The Representational Semantic Conception},
journal = {Philosophy of Science},
volume = {86},
number = {2},
pages = {344-365},
year = {2019},
doi = {10.1086/702029},
}

@book{NicTeh2022,
author = {Nicholas Teh},
title = {Philosophy of Symmetry},
year = {forthcoming},
publisher = {Cambridge, England: Cambridge University Press},
}

@article{Nguyen2017-NGUSRA,
	doi = {10.1086/694003},
	author = {James Nguyen},
	year = {2017},
	number = {5},
	pages = {982--995},
	title = {Scientific Representation and Theoretical Equivalence},
	volume = {84},
	journal = {Philosophy of Science}
}

@article{DeHaro:2021gdv,
    author = "{de Haro}, Sebastian",
    title = "{Noether's Theorems and Energy in General Relativity}",
    eprint = "2103.17160",
    archivePrefix = "arXiv",
    primaryClass = "physics.hist-ph",
    year = "2021"
}

@article{Teitel2021-TEIWTE,
	journal = {Philosophical Studies},
	year = {2021},
	publisher = {Springer Verlag},
	author = {Trevor Teitel},
	number = {12},
	volume = {178},
	pages = {4119--4149},
	title = {What Theoretical Equivalence Could Not Be},
	doi = {10.1007/s11098-021-01639-8}
}

@article{Maudlin2018,
author = {Maudlin, Tim},
year = {2018},
pages = {465},
title = {Ontological Clarity via Canonical Presentation: Electromagnetism and the Aharonov–Bohm Effect},
volume = {20},
journal = {Entropy},
doi = {10.3390/e20060465}
}

@book{healey,
	title = {Gauging What's Real: The Conceptual Foundations of Contemporary Gauge Theories},
	year = {2007},
	editor = {},
	publisher = {Oxford, GB: Oxford University Press},
	author = {Richard Healey}
}

@book{North2021-NORPSA-5,
	publisher = {Oxford: Oxford University Press},
	year = {2021},
	editor = {},
	author = {Jill North},
	title = {Physics, Structure, and Reality}
}

@article{Aharonov:1959fk,
    author = "Aharonov, Y. and Bohm, D.",
    editor = "Taylor, J. C.",
    title = "{Significance of electromagnetic potentials in the quantum theory}",
    doi = "10.1103/PhysRev.115.485",
    journal = "Phys. Rev.",
    volume = "115",
    pages = "485--491",
    year = "1959"
}

@ARTICLE{Vaidman,
       author = {{Vaidman}, Lev},
        title = "{Role of potentials in the Aharonov-Bohm effect}",
      journal = {Phys. Rev. A.},
         year = 2012,
       volume = {86},
       number = {4},
          eid = {040101},
        pages = {040101},
          doi = {10.1103/PhysRevA.86.040101},
archivePrefix = {arXiv},
       eprint = {1110.6169},
 primaryClass = {quant-ph},
       adsurl = {https://ui.adsabs.harvard.edu/abs/2012PhRvA..86d0101V}
}
%\bibliographystyle{dcu}
%\bibliography{boundaries}

\end{document}